%% file: main.tex
\documentclass[11pt,a4paper]{article}
\pdfoutput=1

\usepackage[margin=1in]{geometry}

\usepackage[utf8]{inputenc}
\usepackage[T1]{fontenc}
\usepackage{amsmath,amssymb,mathtools}
\usepackage[separate-uncertainty=true,retain-unity-mantissa=false]{siunitx}
\usepackage{slashed}

\usepackage{graphicx}
\usepackage{booktabs,multicol,multirow}
\usepackage[shortlabels]{enumitem}
\usepackage{hyperref}
\hypersetup{
  colorlinks   = true, 
  urlcolor     = blue, 
  linkcolor    = blue, 
  citecolor   = blue 
}
\usepackage{xspace}
\usepackage[dvipsnames,table]{xcolor}
\usepackage{soul}
\usepackage{subcaption}

\usepackage[frozencache,cachedir=minted-cache]{minted}

\definecolor{bg}{rgb}{0.95,0.95,0.95}
\usepackage{bbold}

\usepackage[capitalize]{cleveref}
\usepackage[numbers,sort&compress]{natbib}
\let\OLDthebibliography\thebibliography
\renewcommand\thebibliography[1]{
  \OLDthebibliography{#1}
  \setlength{\parskip}{2pt}
  \setlength{\itemsep}{0pt plus 0.3ex}
}

\input{abbr.tex}

\begin{document}
\thispagestyle{empty}
\def\thefootnote{\fnsymbol{footnote}}

\begin{flushright}
\texttt{DESY-26-011}\\
\texttt{IFT–UAM/CSIC-25-072}
\end{flushright}
\vspace{.5cm}
\begin{center}
{\large{\huge{\bf
HiggsTools for LHC Run 3 and Beyond
}}}
\\
\vspace{3em}
 {
Henning~Bahl$^{1}$\footnotetext[0]{bahl@thphys.uni-heidelberg.de, 
thomas.biekoetter@desy.de, sven.heinemeyer@cern.ch,\\
\mbox{}\hspace{6mm}kateryna.radchenko@desy.de,
georg.weiglein@desy.de},
Thomas~Biek\"otter$^{2}$,
Sven~Heinemeyer$^{2}$,\\[.3em]
Kateryna~Radchenko~Serdula$^{2}$,
Georg~Weiglein$^{3,4}$
 }\\[2em]
 {\sl $^1$ Institut für Theoretische Physik, Universität Heidelberg, Philosophenweg 16,\\ 61920 Heidelberg, Germany}\\[0.2em]
 {\sl $^2$ Instituto de F\'isica Te\'orica UAM/CSIC,
Calle Nicolás Cabrera 13-15,\\
Cantoblanco, 28049, Madrid, Spain}\\[0.2em]
 {\sl $^3$ Deutsches Elektronen-Synchrotron DESY, Notkestr.~85,\\[0.2em] 22607 Hamburg, Germany}\\[0.2em]
 {\sl $^{4}$ II.\  Institut f\"ur  Theoretische  Physik, Universit\"at  Hamburg, Luruper Chaussee 149,\\[0.2em] 22761 Hamburg, Germany}
\def\thefootnote{\arabic{footnote}}
\setcounter{page}{0}
\setcounter{footnote}{0}
\end{center}
\vspace{2ex}

\begin{abstract}

\HiTo, including the subpackages \HiPr, \HiBo, and \HiSi, is a toolbox for Beyond-the-SM (BSM) scalar phenomenology at the LHC. 
It provides \SH{BSM} model predictions, tests the model against experimental limits from searches for BSM scalars and derives constraints from 
\GW{the measurements of the properties of the} 
discovered Higgs boson. We present a variety of improvements
to 
the \HiTo framework, preparing it for the results of LHC Run~3 and the HL-LHC. \GW{\HiPr} now provides additional cross-section predictions for \GW{centre}-of-mass energies of $13.6$~and~$14\tev$.~Moreover, it now includes \SH{cross-section} predictions for 
resonant and non-resonant Higgs-boson pair production. \GW{For} \HiBo, we 
\GW{describe}
the recasting
of searches using multi-top final states, \GW{explain their}
implementation and \GW{highlight the impact of the experimental sensitivity 
of those results}.
Furthermore, we 
\GW{discuss}
the implementation of coupling-dependent
limits on non-resonant Higgs boson pair production,
as well as the improved handling of
searches conducted prior to the Higgs boson discovery.
\GW{For \HiSi}
we describe several improvements 
\GW{for} the case of scalars with mass uncertainties.

\end{abstract}
\setcounter{footnote}{0}
\renewcommand{\thefootnote}{\arabic{footnote}}

\newpage
{\hypersetup{linkcolor=black}\tableofcontents}
\newpage


\pagestyle{plain}
  
\section{Introduction}
\label{sec:intro}

The discovery of a Higgs boson with a mass of
\TB{about} $125\gev$ at the
LHC~\cite{ATLAS:2012yve,CMS:2012qbp} marks an
important milestone in the quest to unravel the nature of electroweak
symmetry breaking (EWSB).
The further investigation of EWSB constitutes one
of the main tasks of the LHC physics \GW{programme}.  
This comprises both, 
the precise determination of the properties of the
\GW{detected} Higgs boson,
as well as the search for additional scalar bosons.

Many models beyond the Standard Model (BSM) have an extended \TB{Higgs} 
sector
w.r.t.\ the simplest realisation as contained in the Standard Model (SM),
thus predicting additional scalar particles.
Well-studied examples are the extension of the SM Higgs
sector by additional $SU(2)_L$ singlets, and/or doublets, and/or also higher
representations. The LHC searches carried out so far have not yet led to a
discovery of additional \TB{Higgs} bosons. 
Correspondingly, the \GW{limits from those} searches
provide constraints on the parameter space of
BSM models with extended scalar sectors. 

Similarly, also the measurements of the properties of the
125\gev 
Higgs boson have not yet 
\GW{resulted in}
any conclusive deviation from the SM
predictions.
It is thus
also often called the ``SM-like Higgs boson''.
Correspondingly, also these measurements yield constraints on
the parameter space of BSM models,  which --- via mixing, and/or via
quantum corrections --- may 
predict modifications of the
couplings of the
125\gev Higgs boson 
w.r.t.\ the corresponding SM predictions.

Consequently, every BSM model
extending 
the scalar sector of the SM ---
either by adding new BSM scalars to the SM
that can be produced at colliders, and/or by 
modifying the 
\GW{properties of the 125\gev}
Higgs boson ---
should be tested against all the available experimental data collected
at the LHC and other colliders.
A parameter point of a BSM theory
can only be considered viable if it predicts a
SM-like Higgs boson in agreement with the LHC rate measurements, and if
the \GW{predictions for} (possible) additional Higgs bosons are in agreement with all the
existing search bounds. 
Due  to the large number of available searches and measurements, checking the
consistency of a BSM parameter point against all these experimental results is
either imprecise or very cumbersome without the development of dedicated
computer tools to facilitate this task. 

The
codes \HiBo~\cite{Bechtle:2008jh,Bechtle:2011sb,Bechtle:2013wla,Bechtle:2015pma,Bechtle:2020pkv,Bahl:2022igd}
(see also \ccite{Bahl:2021yhk})
and \HiSi~\cite{Bechtle:2013xfa,Bechtle:2020uwn,Bahl:2022igd} have been
developed over the last \GW{decades} to provide such a tool.
\GW{With} \HiBo\ 
\GW{the predictions of}
BSM Higgs sectors 
\GW{can be tested by confronting them with the}
exclusion
limits from searches for new scalar bosons; \HiSi\ 
\GW{can be used}
to check
the compatibility of the model with the LHC rate measurements of the
Higgs boson at $125 \gev$ \GW{as well as with the measurement of the mass of the detected particle}.
In the latest update~\cite{Bahl:2022igd} a
third software component called
\HiPr was presented,
designed to
assist and simplify the computation of the
theory predictions for the production and decay rates
of scalar bosons (both 
for the $125\gev$ Higgs boson 
and 
for additional BSM Higgs bosons) from the model input.
All three tools are now integrated into the main code \HiTo, written in
modern \cpp. The setup allows for an
easy implementation of new \GW{search limits} 
and measurements and provides
simple-to-use \cpp, \py, and \mat interfaces.

In this paper we present the latest version, \texttt{\HiTo-1.3},
containing \texttt{\HiBo-6.1}, \texttt{\HiSi-3.1} and \texttt{\HiPr-2.0},
featuring important updates w.r.t.\ the 
\GW{code}
\TB{presented in} \ccite{Bahl:2022igd}.
For Run~3, the LHC is operating at
a higher \GW{centre}-of-mass energy of $\sqrt{s} = 13.6\tev$.
\HiPr now supports cross-section 
\GW{predictions} not
not only for $\sqrt{s} = 8\tev$ (as part of Run~1)
and $\sqrt{s} = 13\tev$ (for Run~2), but also for
$13.6\tev$ and $14\tev$, the projected \GW{centre}-of-mass
energy of the HL-LHC.
Non-resonant as well as resonant di-Higgs production 
\GW{processes have become a new}
experimental focus over the last years.
Consequently, \HiPr now also provides the
prediction of cross-sections
for the production of two SM-like Higgs
bosons, with or without a BSM
contribution from a heavy Higgs-boson resonance
decaying into \GW{a pair of SM-like Higgs bosons}.
The relevant couplings, e.g.\ the trilinear
self-coupling
of the SM-like Higgs boson, can be varied freely to accommodate possible BSM
\GW{effects}. This facilitates the
tests against the corresponding experimental results,
where \HiBo was updated by incorporating the latest
experimental analyses published by ATLAS and CMS.
Another type of limits that became available over the last years is 
based on 
\TB{searches for spin-0 resonances in}
final states including up to four top
quarks~\cite{CMS:2019rvj,ATLAS:2024jja}.
The inclusion of these limits into \HiBo
(together with information about their derivation) is described in
detail. Also \HiSi was updated, in particular the treatment of theory
uncertainties in the Higgs-boson mass predictions,
as well as the
handling of Higgs-boson mass predictions that (within their
uncertainties) deviate from $125 \gev$,
which concerns models in which the masses are not free \GW{input} parameters
but derived from other model parameters
(e.g.~supersymmetric theories~\cite{Slavich:2020zjv}).

The paper is structured as follows. In \cref{sec:hp} we present the new
features of \HiPr. The \HiBo updates are described in \cref{sec:hb},
together with several phenomenological analyses \KR{that demonstrate their application and relevance.}
The improvements in \HiSi can be found in \cref{sec:hs}, where we 
also include a 
\GW{guidance for how to appropriately use}
the $\chi^2$
provided by \HiSi. 
Our conclusions are 
\KR{presented}
in \cref{sec:conclusions}. Additional details on the implementation of the four-top limit \GW{are} provided in \cref{app:HB}\TB{, while 
the implementation of coupling-dependent cross-section
limits from searches for Higgs-boson pair production
\GW{is described}
in \cref{app:kala_dep_limits}}.


\section{\texorpdfstring{\HiPr}{HiggsPredictions}}
\label{sec:hp}

\HiPr provides cross-section predictions for various processes and \GW{centre}-of-mass energies, as well as corresponding branching ratio predictions.


\subsection{Cross-section predictions for 13.6 and 14~TeV}
\label{sec:newXS}

The first major improvement is that most of the provided tabulated cross-section predictions are now available
for $\sqrt{s} = 13.6\tev$ and $\sqrt{s} = 14\tev$ in addition to the previously available numbers for $\sqrt{s} = 8\tev$ and $\sqrt{s} = 13\tev$. This will be essential for incorporating experimental results from LHC Run-3 and beyond into the 
\HiTo framework. 

\begin{table}[htbp!]\centering
  \begin{tabular}{cccc}
  \hline
  prod.\ channel             & coupling dep.                 & mass range [GeV]                       & source                                    \\
  \hline         
  $gg\phi$                        & $c_t, \tilde c_t, c_b, \tilde c_b$  & $10-3000$                              & \texttt{SusHi} \\
  $b\bar b\phi$                        & $c_b, \tilde c_b$                   & $10-3000$                              & resc.\ of SM result \\
  VBF                          & $c_Z, c_W$                          & $1-3000$                               & \texttt{MadGraph} \\
  $t\bar t \phi$                  & $c_t,\tilde c_t$                    & $25-1000$                              & \texttt{MadGraph}                         \\
  $t\phi$ ($t$ channel)           & $c_t,\tilde c_t, c_W$               & $25-1000$                              & \texttt{MadGraph}                         \\
  $tW\phi$                        & $c_t,\tilde c_t, c_W$               & $25-1000$                              & \texttt{MadGraph}                         \\
  $W\phi$                         & $c_W, c_t$                          & $1-3000$                               & \texttt{vh@nnlo}                          \\
  $qq\to Z\phi$                   & $c_Z, c_t$                          & $1-5000$                               & \texttt{vh@nnlo}                          \\
  $gg\to Z\phi$                   & $c_t,c_b,c_Z,\tilde c_t,\tilde c_b$ & $1-5000$                               & \texttt{vh@nnlo}                          \\
  $b\bar b \to Z\phi$             & $c_b$                               & $1-5000$                               & \texttt{vh@nnlo}    \\
  $gg \to hh$& $c_t, \kappa_{\lambda}, c_3$        & $251-3000$
                          & \texttt{anyHH}   \\
  \hline
  \end{tabular}
  \caption{Overview of 13.6 and 14 TeV cross-section predictions available in \HiPr for scalars with a non-SM-like coupling structure. $\phi$ denotes a generic (BSM) Higgs boson\KR{\st{s}}, while $h$ denotes the SM-like Higgs boson discovered at the LHC. $c_3$ is the product of BSM couplings in the resonant di-Higgs production \GW{process}, $c_3 \equiv c_t^H\times \lambda_{hhH}$, see text for details.}
  \label{tab:tabulated_XS}
\end{table}

First of all, this concerns cross-section predictions for scalars with a non-SM-like coupling structure.~We provide an overview of these in
\cref{tab:tabulated_XS}. Here, $\phi$ denotes a generic (BSM) Higgs boson\KR{\st{s}}, while $h$ here and in the following 
denotes the SM-like Higgs boson discovered at 
the LHC; $c_Z$ and $c_W$ are used to denote the couplings of the scalar to $Z$ and $W$ bosons normalised to the respective coupling of the 
SM Higgs boson (see \ccite{Bechtle:2020pkv} for more details). Similarly, we use $c_q$ and $\tilde c_q$ to respectively denote the \cp-even and \cp-odd 
components of the Yukawa coupling of
$\phi$ to the quark $q$, which
are both normalised to the magnitude of the
respective \cp-even
SM Yukawa coupling (see \ccite{Bechtle:2020pkv} for more
details).\footnote{\HiPr does not consider
flavour-changing $\phi q \bar q$ couplings.}
The coefficient
\begin{align}
    \kappa_\lambda \equiv \TB{\frac{\lambda_{hhh}}{\lambda_{hhh}^{\mathrm{SM},(0)}}}
    \label{eq:kaladef}
\end{align}
is defined as the ratio between the trilinear self-coupling
of the scalar $h$ in the respective BSM theory,
\TB{$\lambda_{hhh}$},
and the leading-order~(LO) Higgs boson self-coupling
predicted in the~SM, \TB{$\lambda_{hhh}^{\rm SM,(0)}$}.
Finally,
$c_3$ denotes the product of the
couplings of the heavier BSM Higgs boson
involved in resonant Higgs-boson pair production, 
normalised as explained in \cref{sec:hhXS}. 
We have derived the gluon fusion 
cross-sections $\sigma(gg\phi)$ using \texttt{SusHi~1.7.0}~\cite{Harlander:2012pb,Harlander:2016hcx}. For the VBF channel, we employ \texttt{MadGraph5\_aMC@NLO~2.8.2}~\cite{Alwall:2014hca}. Also for the top-associated Higgs boson production channels, \texttt{MadGraph} is used,
and we extended the mass range from $1\tev$ to $3\tev$ in addition to the new centre-of-mass energy values.
For vector-boson associated production, we employed \texttt{vh\@nnlo 2.1}~\cite{Brein:2012ne,Harlander:2018yio}, with cross-checks against \texttt{MadGraph}~\cite{Bahl:2020wee}. Additionally, we provide cross-sections for Higgs 
boson pair production via gluon fusion, 
derived with \texttt{anyHH}~\cite{Bahl:2023eau, anyHH},
which provides the LO $gg\to hh$ cross-section
for generic SM extensions based on a user-defined \texttt{UFO} model file, and which \GW{is} rescaled to incorporate full NLO QCD results for the resonant case and to 
NNLO
for the non-resonant one~\cite{Borowka:2016ehy,Borowka:2016ypz,Baglio:2018lrj,Davies:2019dfy,Baglio:2020wgt,deFlorian:2013jea,Grigo:2014jma,Davies:2021kex,Baglio:2020ini,Grazzini:2018bsd,LHCHWG4twiki}, as described in detail in \refse{sec:hhXS}.

As explained in \ccite{Bahl:2022igd}, these BSM predictions are then scaled by the ratio of the LHC Higgs working group (LHCHWG) recommendation at that mass value divided by the cross-section prediction for a SM-like scalar.
For this, we use the newly provided recommendations of the LHCHWG for $\sqrt{s} = 13.6\tev$ and $\sqrt{s} = 14\tev$~\cite{LHCHWG4twiki}.\footnote{We have approximated the numbers for $t\bar t H$ production by multiplying the $13\tev$ numbers by a mass-dependent energy rescaling factor derived by taking the ratio of the predictions for $\sqrt{s} = 13.6,14\tev$ and $\sqrt{s} = 13\tev$ obtained using \texttt{MadGraph}. These numbers have by now become the official recommendation of the LHCHWG.}

Moreover, we provide predictions including electroweak NLO corrections for a SM-like scalar in the mass window between 120 and 130 GeV based on the LHCHWG recommendation published in \ccite{Karlberg:2024zxx}.
Here, we 
fill up missing mass values by multiplying the $13\tev$ numbers with an energy rescaling factor, which we in this case assume to be constant in the considered mass window.


\subsection{Interpolated reference model}
\label{sec:SMHiggsInterp}

\begin{figure}[hptb!]
    \centering
    \includegraphics[width=0.7\textwidth]{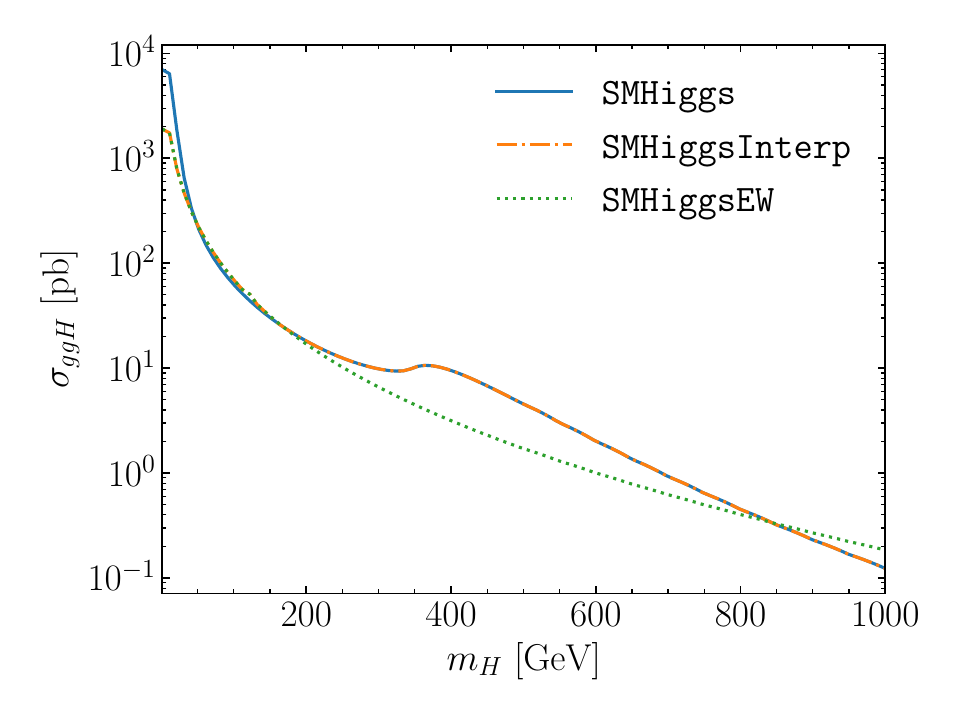}
    \caption{Comparison of \GW{the} gluon fusion cross-section \GW{predictions} for the various reference models.}
    \label{fig:ref_models}
\end{figure}

As explained above, \HiPr uses SM reference cross-sections
to approximately scale BSM cross-section predictions to higher-order accuracy. For the gluon fusion cross-section, different precision levels are implemented in the form of reference models. These are:
\begin{itemize}
    \item \texttt{SMHiggsEW}: N$^3$LO QCD corrections in the infinite top quark mass limit and including NLO electroweak corrections.
    \item \texttt{SMHiggs}: NNLO QCD corrections for a finite top quark mass.
\end{itemize}
While \texttt{SMHiggsEW} is more precise for scalar masses $m_S < m_t$, \texttt{SMHiggs} is more accurate for $m_S > m_t$~\cite{Aglietti:2004nj,Anastasiou:2014lda,Anastasiou:2015yha,Anastasiou:2016cez,LHCHiggsCrossSectionWorkingGroup:2016ypw}. As shown in \ccite{Bahl:2022igd}, this difference is relevant for phenomenological applications. 
In the previous version of \HiTo, the reference model had to be
specified by the user, often leading to inaccurate choices. 
To avoid these issues, we \GW{have introduced} the \texttt{SMHiggsInterp} reference model which interpolates between the \texttt{SMHiggsEW} reference model (for low masses) and the \texttt{SMHiggs} reference model (for high masses). The resulting predictions for the gluon fusion cross-section are shown in \cref{fig:ref_models}. The \texttt{SMHiggsInterp} reference model is the new default reference model.

\subsection{Predictions for \texorpdfstring{\boldmath{$hh$}}{h125h125} production}
\label{sec:hhXS}

In addition to incorporating new LHC cross-section predictions
for
centre-of-mass energies 
of 13.6 and $14\tev$, as well as the new 
interpolated reference model \texttt{SMHiggsInterp},
\HiPr now also provides 
predictions for the pair production cross-section of two 125~GeV Higgs bosons $h$.
We provide \GW{predictions} for resonant, non-resonant, and combined resonant plus non-resonant Higgs boson pair production. The current implementation supports scenarios involving a single BSM resonance decaying into a $h$-pair.

\begin{figure}[!hbt]
    \centering
    \includegraphics[width=0.9\textwidth]{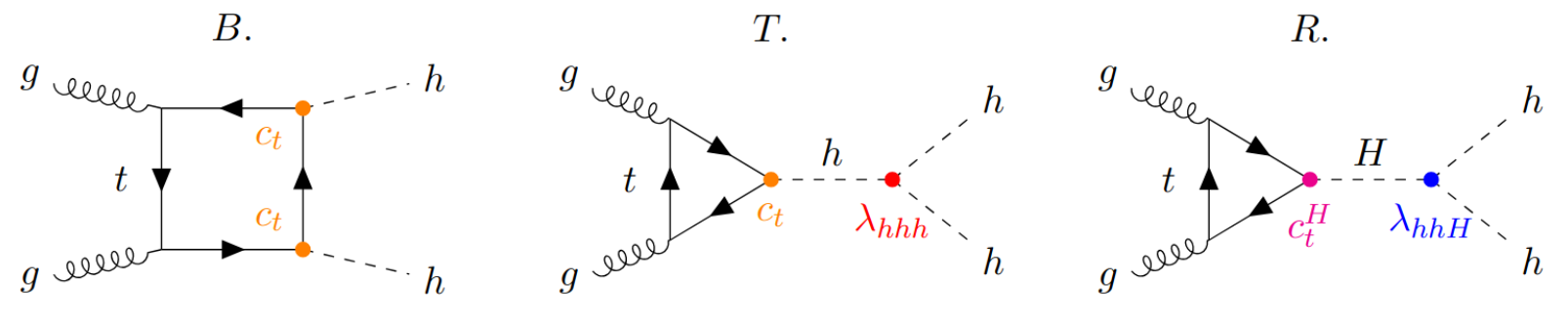}
    \caption{Leading order diagrams for $h$-pair production via gluon fusion.}
    \label{fig:hhdiags}
\end{figure}

The non-resonant production cross-section is obtained using the recommendation of the LHCHWG~\cite{LHCHWG4twiki,Amoroso:2020lgh,Heinrich:2022idm,Bagnaschi:2023rbx}, which we modify to account for \GW{a possible} 
deviation in the 
\cp-even component of the
$h$ top quark Yukawa coupling,~$c_t$. For the case of $\sqrt{s}=13\tev$, 
the cross-section for $c_t = 1$ can be approximated by
a quadratic function of $\kappa_{\lambda}$,\footnote{We note that \HiPr returns the cross-section in picobarn.}
\begin{equation}
    \sigma_{\mathrm{no\, res}}^{13\,\text{TeV}} (\kappa_{\lambda}) {\mathrm{/fb}} = 68.5624-48.3673\, \kappa_{\lambda}+10.5635\, \kappa_{\lambda}^2 \;,
    \label{eq:lhchwg}
\end{equation}
obtained for a Higgs \TB{boson} mass
of $m_h = 125$~GeV with the renormalisation and factorisation
scales set to 
$\mu_0 = m_{\KR{h}\KR{h}}/2$. 
To account for the impact of $c_t \neq 1$, we modify
this equation 
to
\begin{equation}
    \sigma_{\mathrm{no\, res}}^{13\,\text{TeV}} (c_t, \kappa_{\lambda}){\mathrm{/fb}} = 68.5624\, c_t^4 -48.3673\, c_t^3\, \kappa_{\lambda}+10.5635\, c_t^2\, \kappa_{\lambda}^2\;.
    \label{eq:sigma_nores_HT}
\end{equation}
The additional factors of $c_t$ were incorporated according
to the number of $ht \bar t$-vertices contained
in the different diagrams contributing
to the cross-section at leading order:
The first term on the right-hand side accounts for the 
\GW{squared}
contribution of the box diagram with a top quark loop and has no contribution from the trilinear $h$ self-coupling
proportional to $\kappa_\lambda$
(see 
\GW{left} 
diagram in \cref{fig:hhdiags}).
The third term 
stems from the \GW{squared} triangle diagram with the exchange of $h$ in the propagator (see 
\GW{middle}
diagram of \cref{fig:hhdiags}), proportional to the second
power of both $\kappa_\lambda$ and $c_t$.
The second term comes from the
destructive interference of these two diagrams and is
therefore negative.
The LHCHWG also provides the coefficients in the parametric form as in \cref{eq:lhchwg} for a centre-of-mass energy of 13.6\tev, which is also implemented in \HiPr. For 14\tev we have rescaled the result at LO obtained with \texttt{anyHH} with a $K$-factor derived from the LHCHWG recommendations at 13\tev and 13.6\tev. 

The functional dependence of $\sigma_{\mathrm{no\,res}}$ on $c_t$ has also been investigated in an effective field theory (EFT) approach in \ccite{Alasfar:2023xpc}, where the dependence on 
further effective couplings
in addition to $c_t$ and $\kappa_\lambda$
has been included.
A fit-formula for the cross-section as a function of
Wilson coefficients in the Higgs~EFT~(HEFT) is presented
(see Eq.~(4.1) in \ccite{Alasfar:2023xpc})
that can be mapped onto the fit-formula of the LHCHWG
given in \cref{eq:lhchwg}.
However, the coefficients of the formula
presented in \ccite{Alasfar:2023xpc} for $\kappa_{\lambda}$ floating and the other couplings fixed to their SM values differ from the recommendations of the LHCHWG, as only the NLO corrections are taken into account. 
\GW{In contrast}, the LHCHWG recommendations are rescaled to the NNLO correction in the $\text{FT}_\text{approx}$ approximation for $\kappa_{\lambda} = 1$~\cite{Grazzini:2018bsd}. 
Therefore, we chose to implement 
the functional dependence on the coupling $c_t$
by modifying the LHCHWG recommendation
as shown in \cref{eq:sigma_nores_HT}.

The function 
returning
the non-resonant $hh$ production cross-section can be called in \texttt{python} 
by running:
\begin{minted}[bgcolor=bg]{python}
import Higgs.predictions as HP

pred = HP.Predictions() # create the model predictions object

# compute only the non-resonant contribution
ggHH_XS_nores = HP.EffectiveCouplingCxns.ggHHnores(
    'LHC13',
    lam=1,
    tt=1)
\end{minted}
This yields the non-resonant di-Higgs production cross-section at 13\tev for $\kappa_\lambda =1$ (controlled via
the argument \texttt{lam}) and $c_t = 1$
(controlled via the argument \texttt{tt})
in units of picobarns, in this case $\sigma_{\mathrm{no\,res}}^{\mathrm{13TeV}} = 30.76$~fb.
This is the default recommendation, although if preferred, the HEFT formula can be
used to calculate the non-resonant
pair production cross-section by calling:
\begin{minted}[bgcolor=bg]{python}
ggHH_XS_nores_HEFT = HP.EffectiveCouplingCxns.hhggh(
    'LHC13', 
    mH=125, 
    lam=1,
    tt=1,
    bb=1)
\end{minted}
where additionally the mass of the SM-like Higgs
boson, \texttt{mH,} and the bottom Yukawa coupling modifier, \texttt{bb}, 
need to be provided.
This yields a cross-section of $\sigma_{\mathrm{HEFT}}^{13\text{TeV}} = 31.66$ fb.
It should be noted that these two functions only
compute the respective cross-sections. However,
in order to apply the experimental limits, they still
need to be assigned to the Higgs boson that 
\GW{corresponds to}
the SM-like Higgs boson at 125\gev,
as is explained in the following.

\begin{figure}[htb!]
    \centering
    \includegraphics[width=0.65\textwidth]{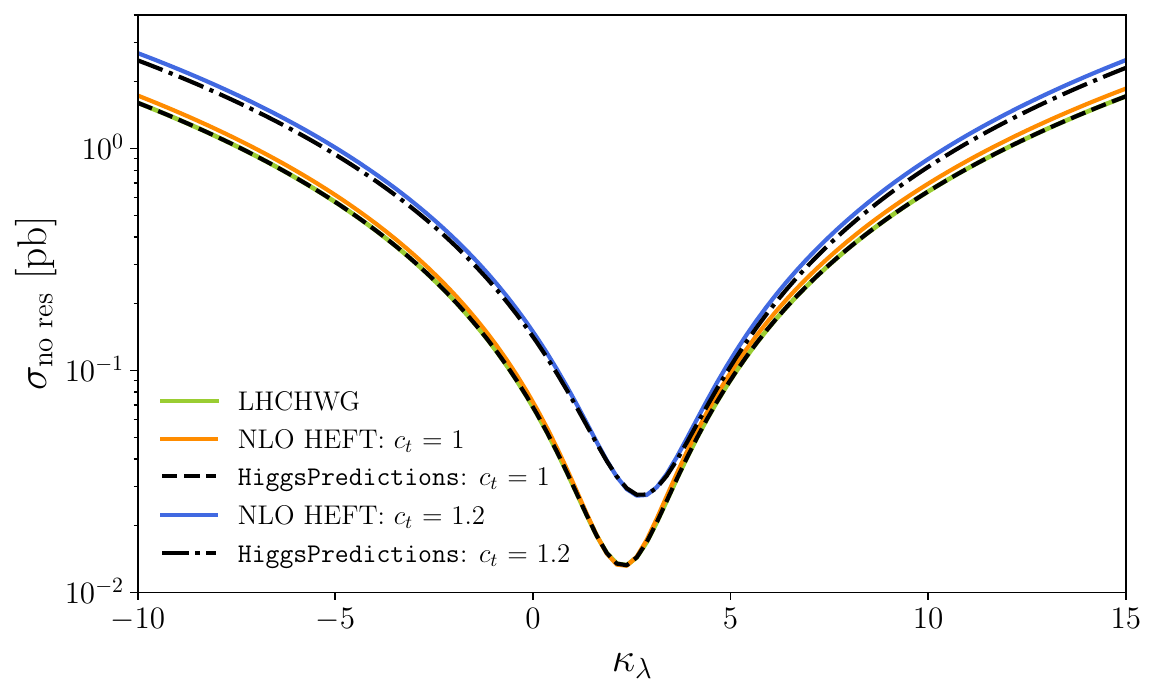}
    \caption{Prediction of the non-resonant Higgs \TB{boson} pair production cross-section in terms of $\kappa_\lambda$. In black, the prediction implemented in \HiPr is shown for different values of $c_t$ ($c_t=1$ for dashed and $c_t=1.2$ for dot-dashed); in green, the LHCHWG recommendation; and, in orange (blue), the prediction of \ccite{Alasfar:2023xpc} for $c_t = 1$ $(c_t=1.2)$ \GW{is shown}.}
    \label{fig:pred_nores}
\end{figure}

In \cref{fig:pred_nores}, we show the difference between the two 
types of predictions contained in \HiPr. The LHCHWG recommendation is only given for $c_t = 1$
and shown with the green line. It 
coincides with the default implementation in \HiTo
with $c_t = 1$ indicated with the 
black dashed line. In orange, we show the 
\GW{prediction} based on \ccite{Alasfar:2023xpc} for $c_t = 1$, which 
\GW{yields} a cross-section slightly above the one
of the \HiTo default implementation. For $c_t = 1.2$ the \HiTo\ default implementation 
(black dashed-dotted line)
shows good agreement 
\GW{with the result of \ccite{Alasfar:2023xpc}. It can be seen that the 
\HiTo prediction}
is slightly below the result of \ccite{Alasfar:2023xpc}, as expected due to the difference in the numerical coefficients in \cref{eq:sigma_nores_HT} w.r.t.\ the ones in the HEFT formula.

Moreover, \HiTo now
also \GW{provides} predictions for the 
Higgs boson pair production cross-section including
the contribution from the resonant production
via the decay of an additional \cp-even BSM 
scalar, denoted as $H$ in the following.
The diagrams contributing to the full cross-section are shown in \reffi{fig:hhdiags}.~At the leading order, this process includes the contribution of the 
resonant diagram~($R$) with $H$ in the
$s$-channel, which is proportional to the product of BSM couplings $c_3 \equiv \lambda_{hhH} \times c_t^H$
, where $c_t^H$ is the ratio between the
strength of the $H t \bar t$ Yukawa coupling in the BSM theory and the one of the SM. 
\GW{Here} $\lambda_{hhH}$ 
\GW{is} the dimensionless coupling between two $h$ Higgs bosons and the 
BSM Higgs boson $H$, such that the Feynman rule is given by 
\begin{equation}
    - i\,2\,v\,\lambda_{hhH}\, ,
    \label{eq:lahhH}
\end{equation}
where $v \approx 246.2\gev$ is the vacuum expectation
value of the Higgs field in the SM.
The process furthermore includes the box ($B$) and the
triangle ($T$) diagrams from the non-resonant
production.


\TB{
Before describing our implementation of the full Higgs-boson pair-production cross section, which
includes the non-resonant, resonant, and interference 
contributions, we first focus on the purely resonant part.
Resonant Higgs-boson pair production could already be
obtained with the previous version of \HiTo using
the narrow-width approximation. There, one sets 
explicit values for the $H$ production cross section 
and for the branching ratio $H \to hh$,
using the functions \texttt{H.setCxn()} and \texttt{H.setDecayWidth()}/\texttt{H.setBr()}, respectively (see Sec.~4.2 of Ref.~\cite{Bahl:2022igd}).
In the present implementation we add a new function,
\texttt{ggHHres()}, which instead computes the 
resonant contribution directly from the effective
coupling input.
\KR{For the resonant contribution, only the coupling $c_3$, which already includes the trilinear coupling $\lambda_{hhH}$, together with the mass $m_H$ and the total width $\Gamma_H$, is required. The additional couplings $c_t$ and $\kappa_\lambda$ listed in \cref{tab:tabulated_XS}  enter only the non-resonant and full production.}
This follows the same input scheme used
for the computation of the non-resonant and full 
$hh$ production, see the discussion below.
The advantage of this new function is that the user
no longer has to compute $\sigma(gg \to H)$ and $
\mathrm{BR}(H \to hh)$ separately.
Instead, the resonant-only contribution is obtained
from the very same input that is needed for the
full process.
Like the earlier approach, the new \HiPr
implementation relies on the narrow-width
approximation, splitting the process into gluon-fusion production of the heavy resonance $H$
followed by its decay into a pair of $h$,
\begin{equation}
    \sigma(gg \to H \to hh)
    \,\approx\,
    \sigma(gg \to H)\times \mathrm{BR}(H \to hh) \, .
    \label{eq:sigmaresonant}
\end{equation}
The total cross section $\sigma(gg \to H)$ is taken from the existing \HiPr implementation of $ggH$
production, which uses tabulated results from
\texttt{SusHi 1.7.0}~\cite{Harlander:2012pb,Harlander:2016hcx},
considering only the CP-even component of the
top-quark Yukawa coupling $c_t$.
The cross section is
rescaled to the LHCHWG recommendations
and therefore 
includes the N$^3$LO corrections to the process
(see \cref{sec:newXS}).
The branching ratio is computed as
\begin{equation}
    \mathrm{BR}(H \to hh) = \frac{\lambda_{hhH}^2}{32\pi m_H} \sqrt{1 - \frac{4 m_h^2}{m_H^2}} \frac{1}{\Gamma_H} \, ,
    \label{eq:BRresonant}
\end{equation}
which assumes that loop corrections to this decay are either negligible or can be absorbed into an effective, loop-corrected coupling that the user supplies as the value of $\lambda_{hhH}^2$.}
\TB{The cross section computed in this way (\texttt{xs})
is returned by the function but is \textit{not} 
stored automatically as a property of the particle 
object. To have the experimental limits applied,
the user therefore has to assign it explicitly with
\texttt{setCxn('LHC13', 'pair', xs)}, as shown in
the code snippet below.
Alternatively, as in the previous version of
\HiTo, the user can provide the branching ratio for
the decay $H \to hh$ and the
resonant $H$ production cross section separately. In this case no pair-production cross section needs
to be assigned explicitly, because \HiBo automatically 
constructs it in the narrow-width approximation
from the product of the two inputs.}

As an example, the 
Higgs \TB{boson} pair production
cross-section 
\TB{resulting from a new spin-0 resonance} with a mass of $400 \gev$ 
and a width of $4 \gev$
\TB{decaying into $h$-pairs} can be 
\TB{obtained} in python by typing
\begin{minted}[bgcolor=bg]{python}
import Higgs.predictions as HP
pred = HP.Predictions() # create the model predictions object

# set mass and width of the resonant particle
mH = 400
widthH = 4

# add the resonant particle with the above properties
H = pred.addParticle(HP.BsmParticle("H", "neutral", "even"))
H.setMass(mH)
H.setTotalWidth(widthH)

# compute only the resonant contribution
ggHH_XS_res = HP.EffectiveCouplingCxns.ggHHres(
    'LHC13',
    mH=mH, 
    totalWidth=widthH, 
    c3=-0.1)
\end{minted}
where the symbol \texttt{c3} refers to the product
$c_3$. 
\TB{We stress that to compute the value of
\texttt{ggHH\_XS\_res}, the branching ratio
is computed according to \cref{eq:BRresonant} using the
provided effective couplings. If the user has
previously assigned a value of
$\mathrm{BR}(H \to h h)$, it is not
used here when calling the function
\texttt{ggHHres}.}

\GW{While up to now in our examples for calling \HiPr 
we have focussed on the pure non-resonant
$h$-pair production and the pure resonant
$h$-pair production via the decay of a heavier
BSM spin-0 resonance $H$, 
in general both contributions
and their interference 
should be taken into account.
We now describe the procedure that we have used in order to provide 
an accurate prediction for the 
\TB{$h$-}pair production cross-section containing the resonant, non-resonant and all relevant interference contributions.} 
We implement a prediction for the
$h$-pair production cross-section
as a function of the couplings $c_t$, $\kappa_\lambda$
and $c_3$, the mass $m_H$ of the BSM resonance
and its total width $\Gamma_H$ by adding
the non-resonant, resonant and interference terms as
\begin{equation}
\begin{split}
    \sigma_{\mathrm{tot}} (c_t, \kappa_{\lambda}, c_3, m_H, \Gamma_H) =&\, c_t^2\kappa_{\lambda}^2\, K_T \mathcal{A_{\mathrm{T}}}\, + c_t^4\, K_B\mathcal{A_{\mathrm{B}}}\, + c_t^3\kappa_{\lambda}\, K_{TB} \mathcal{A_{\mathrm{TB}}}\, + \\
    & c_3^2\,K_R (m_H)\mathcal{A_{\mathrm{R}}}(m_H, \Gamma_H) \, + \\
    & c_t \kappa_{\lambda} c_3\, K_{RT} (m_H)\mathcal{A_{\mathrm{RT}}}(m_H, \Gamma_H) \, +\\
    &c_t^2 c_3\, K_{RB} (m_H)\mathcal{A_{\mathrm{RB}}}(m_H, \Gamma_H), 
\end{split}
    \label{eq:sigma_hh_nlo}
\end{equation}
\GW{taking into account $K$-factors for all the contributions, see below.}
Here 
\GW{$\mathcal{A_{\mathrm{T}}} \sim |\mathcal{M}_T|^2$,
$\mathcal{A_{\mathrm{B}}} \sim |\mathcal{M}_B|^2$,}
$\mathcal{A_{\mathrm{R}}} \sim |\mathcal{M}_R|^2$, 
\GW{$\mathcal{A_{\mathrm{TB}}} \sim 2\mathrm{Re}[\mathcal{M}_T\mathcal{M}_B^*]$,}
$\mathcal{A_{\mathrm{RT}}} \sim 2\mathrm{Re}[\mathcal{M}_R\mathcal{M}_T^*]$ and $\mathcal{A_{\mathrm{RB}}} \sim 2\mathrm{Re}[\mathcal{M}_R\mathcal{M}_B^*]$, 
\GW{where $\mathcal{M}_T$, $\mathcal{M}_B$ and $\mathcal{M}_R$ are the scattering amplitudes and 
$\mathcal{A_{\mathrm{T}}}$, 
$\mathcal{A_{\mathrm{B}}}$, 
$\mathcal{A_{\mathrm{R}}}$, 
$\mathcal{A_{\mathrm{TB}}}$,
$\mathcal{A_{\mathrm{RT}}}$ and
$\mathcal{A_{\mathrm{RB}}}$ are understood to be
integrated over the phase space.}
\GW{The products of amplitudes $\mathcal{A_{\mathrm{R}}}$, 
$\mathcal{A_{\mathrm{RT}}}$ and
$\mathcal{A_{\mathrm{RB}}}$}
contain the dependence on the mass and the width
of the BSM resonance which cannot be expressed in a
simple analytic form; the subindices $R$, $B$ and $T$ refer to the diagrams shown in \reffi{fig:hhdiags}. 
\GW{The products of amplitudes}
have been computed
as a function of $m_H$ and $\Gamma_H$ in a 
two-dimensional grid \GW{as described below} 
and are stored in the form of
data tables in \HiPr.
The formula shown in \cref{eq:sigma_hh_nlo}
can be applied 
under the condition that the gluon fusion
production of 
$h$ is only altered by the presence of
one BSM resonance and modifications to the
top-quark Yukawa coupling, 
as long as the top-quark loop dominates over all other contributions. In particular, we do not include the effects of the bottom loop, which contribute at the per-cent-level \GW{in the SM}~\cite{Baglio:2020ini}.
Furthermore, as for $h$, we assume that the BSM resonance $H$ is
dominantly produced via gluon fusion production
and that only the contribution from the top-quark
loop is relevant.

In order to obtain the interference terms \GW{between the resonant and non-resonant contributions}, $\mathcal{A_{\mathrm{RT}}}(m_H, \Gamma_H)$ and $\mathcal{A_{\mathrm{RB}}}(m_H, \Gamma_H)$, we use the framework of \texttt{anyHH}~\cite{Bahl:2023eau,anyHH}.
To this end, we have defined a
minimal UFO 
model file that, in addition to the SM-like
Higgs boson $h$, contains one 
heavy scalar which can only be produced through gluon fusion and 
\GW{decays} to a pair of 125 GeV Higgs-bosons. The 
terms added to 
the SM Lagrangian that will generate such a resonant contribution can be written as 
\begin{equation}
    \mathcal{L} = \frac{1}{2} \partial_{\mu} H \partial^{\mu} H - \frac{1}{2} m_H^2 H^2 - \lambda_{hhH} v hhH - c_t^H \frac{m_t}{v} \bar{t}t H \, ,
\end{equation}
where $\lambda_{hhH}$ and $c_t^H$ are the aforementioned dimensionless parameters. We note that the Lagrangian defined in such a way 
\GW{gives} rise to the Feynman rule in \cref{eq:lahhH}, where a symmetry factor of 2 appears, which needs to be taken into account to map the couplings to a particular model. 



\GW{For the pure non-resonant and the pure resonant contribution, the $K$-factors are obtained from the precise predictions described above. Specifically, we have rescaled}
\KR{the LO QCD prediction obtained from \texttt{anyHH} to the NLO QCD predictions of the LHCHWG and \texttt{SusHi} to obtain the resonant and non-resonant $K$-factors.}
In order to 
approximately include NLO QCD corrections
in the interference terms we have defined
multiplicative $K$-factors $K_{TB}$, $K_{RT}$
and $K_{RB}$
for each value of $(m_H, \Gamma_H)$ 
and for the
three interference terms involved
in the Higgs boson pair production process: the interference between the box and the triangle diagrams
$\mathcal{A_{\mathrm{TB}}}$
(captured by the negative coefficient of \cref{eq:sigma_nores_HT}), the interference between the triangle and the resonant diagram $\mathcal{A_{\mathrm{RT}}}$, and
the interference between the box and the resonant diagram $\mathcal{A_{\mathrm{RB}}}$. 
%
%
%
For the interference terms \GW{between the resonant and non-resonant contributions}, we approximate the $K$-factors 
as
\begin{equation}
    K_{RT} = \sqrt{K_R K_T} \quad \textrm{and} \quad
    K_{RB} = \sqrt{K_R K_B} \, .
    \label{eq:K_fac}
\end{equation} 

The $K$-factors for the non-resonant part are
\TB{constant} numbers, in our case $K_T = 2.989$, $K_B =  2.492$, and the interference of the box and triangle diagram, $K_{TB} = 2.735$. \KR{These numbers are obtained from the ratio of the loop-corrected LHCHWG prediction \SH{and} 
the \texttt{anyHH} LO result for each term of the non-resonant production cross-section.} 
\GW{Applying}
the approximation of \cref{eq:K_fac} 
\GW{also for $K_{TB}$ would have resulted in}
a 1\% deviation 
\GW{from the value given above.}

\begin{figure}[!htb]
    \centering
    \includegraphics[width=0.65\linewidth]{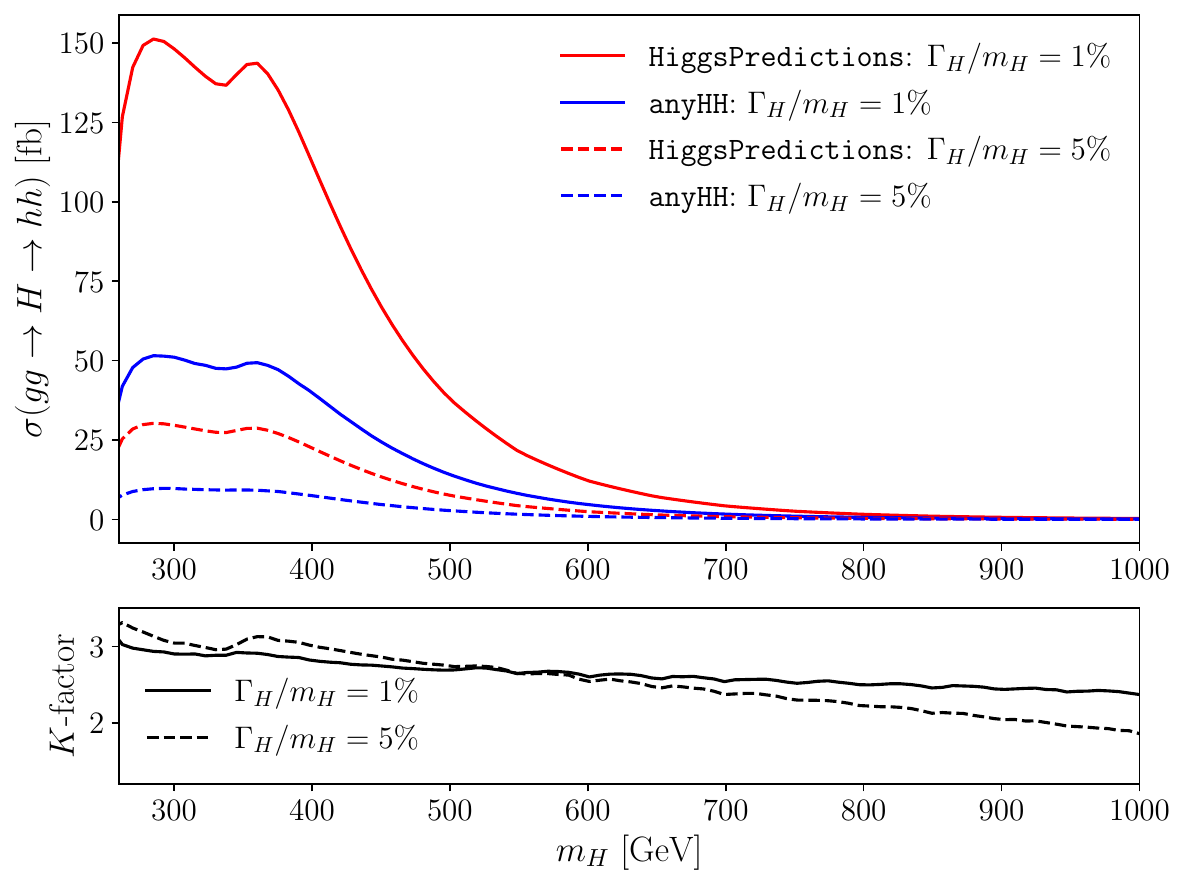}
    \caption{Cross-section and $K$-factors for resonant di-Higgs production computed 
    \GW{using} 
    \HiPr, \GW{based on the} 
    tabulated values for 
    \TB{the production of a single Higgs boson}
    \GW{according to \cref{eq:sigmaresonant} and \cref{eq:BRresonant}, and}
    \texttt{anyHH}.
}
    \label{fig:kfactor_res}
\end{figure}

\KR{The resonant $K$-factor, $K_R(m_H)$, is computed as the ratio of the \GW{\HiPr result, based on
\texttt{SusHi}
(see \cref{eq:sigmaresonant} and \cref{eq:BRresonant}),}
and the \texttt{anyHH} result at each resonant mass value. 
\GW{As a consequence of the off-shell effects implemented in \texttt{anyHH} and the modification of the branching ratio in \cref{eq:BRresonant},}
we find a slight dependence of this ratio on the width of the resonant scalar. 
\GW{This is shown} in \reffi{fig:kfactor_res}, where in the upper panel the prediction of \texttt{anyHH} is displayed in blue for two different values of the total decay width, $\Gamma_H/m_H = 1\%$ and $\Gamma_H/m_H = 5 \%$, and in red the corresponding prediction implemented in \texttt{HiggsPredictions} is shown. 
The lower panel shows the $K$-factor for the two considered width values. 
\GW{Since this $K$-factor shows only a moderate dependence on the width, $\Gamma_H$, for simplicity we}
apply the same $K$-factor for each resonant mass
instead of defining a width-dependent $K$-factor.
\GW{For this purpose we choose the $K$-factor} corresponding to the solid black line of the lower panel (which is close to 3), i.e.\ we}
\KR{define  $K_R(m_H) \equiv K_R(m_H, \Gamma_H = 0.01m_H)$
for the resonant $K$-factor $K_R(m_H)$ 
entering the calculation of the interference terms.}

\KR{We extract the amplitudes of the interference terms $\mathcal{A_{\mathrm{RT}}}$ and $\mathcal{A_{\mathrm{RB}}}$ at LO in QCD from the code \texttt{anyHH} using different coupling combinations in order to 
\GW{determine} each term separately. We apply the $K$-factors using \cref{eq:K_fac}.}
\GW{The different terms in \cref{eq:sigma_hh_nlo} that we have determined as described above are saved in the form of grids in \HiPr, which are linearly interpolated to obtain values at non-grid points.}

An example script 
to set the prediction of the 
di-Higgs production cross-section \GW{such that it contains}
both the non-resonant and the
resonant contribution as well as the interference
effects, as discussed above, is given below:
\begin{minted}[bgcolor=bg]{python}
import Higgs.predictions as HP

pred = HP.Predictions() # create the model predictions object

# set mass and width of the heavy BSM scalar
mH = 400
widthH = 4

# add the BSM scalar with the above properties
H = pred.addParticle(HP.BsmParticle("H", "neutral", "even"))
H.setMass(mH)
H.setTotalWidth(widthH)

# compute the total Higgs-boson pair production cross-section
# specifying the couplings
ggHH_XS = HP.EffectiveCouplingCxns.ggHH(
    'LHC13',
    mH=mH, 
    totalWidth= widthH,
    lam=0.8,
    c3=-0.1,
    tt=(1.2+1j))

# assign the computed value to the internal 'pair' cross-section
# to apply experimental bounds
H.setCxn('LHC13', 'pair', ggHH_XS)

\end{minted}
Here we set the collider to \texttt{LHC13},
introduce a BSM Higgs boson with a mass
of $m_H = 400\gev$ and a width of $\Gamma_H = 4\gev$,
and the relevant couplings are set to
$\kappa_\lambda = 0.8$, 
$c_3 =-0.1$ and $c_t^H =1$. In the final command,
the Higgs boson pair production cross-section
is 
assigned to the BSM scalar $H$ with the value
calculated \KR{via the function \texttt{EffectiveCouplingCxns.ggHH()}} for \KR{the subsequent} application of \KR{ experimental limits with \HiBo}. 

Prior to the implementation of predictions for
Higgs boson pair production cross-sections
developed in this work, the cross-section had to be supplied
manually by the user 
based on an independent cross-section calculation
for the specific model at hand. \KR{In the new implementation, if no Higgs \TB{boson} pair production cross-section is assigned explicitly, the code automatically uses the resonant prediction calculated with the 
\GW{narrow width approximation}, provided that the user has specified the gluon-fusion production cross section of the BSM Higgs boson and its branching ratio to \TB{$hh$}. 
The new implementation provides greater flexibility in assigning Higgs-pair production cross sections for the application of experimental limits. Depending on the desired level of accuracy, the user can compute with \HiPr the full Higgs \TB{boson} pair production cross section -- including the resonant and non-resonant contributions as well as their interference -- using \texttt{EffectiveCouplingCxns.ggHH()}, or only the resonant contribution using \texttt{EffectiveCouplingCxns.ggHHres()}. More generally, any user-provided Higgs \TB{boson} pair production cross-section, with or without interference effects, can be assigned via \texttt{setCxn()}. We note, however, that the currently available experimental limits on resonant Higgs \TB{boson} pair production neglect the non-resonant and interference contributions, although these effects can be significant in general.~\cite{Heinemeyer:2024hxa}}

\begin{figure}[!htb]
    \centering
    \includegraphics[width=0.65\linewidth]{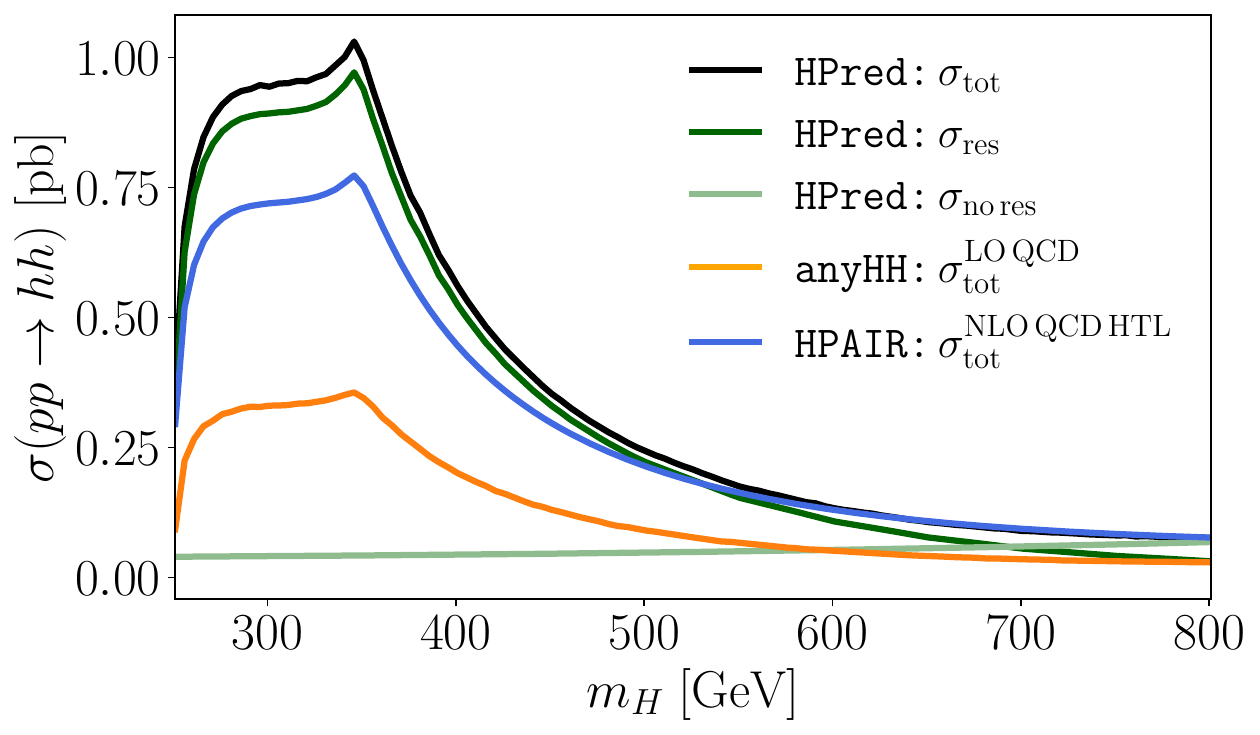}
    \caption{Contributions to the Higgs pair production cross-section in the 2HDM \GW{of type} II (with
    $\tan\beta = 2,\; \cos(\beta-\alpha)= 0.1,\; m_H=m_A=m_{H^{\pm}}=M$) \GW{evaluated} with \HiPr: full cross-section (black), resonant only contribution (dark green) and non-resonant contribution (light green). The LO result in \texttt{anyHH} is shown in orange and the NLO QCD \GW{result} in the heavy top limit of \texttt{HPAIR} in blue. No constraints \GW{are} applied.}
    \label{fig:gghh_thdm}
\end{figure}

In \reffi{fig:gghh_thdm} we show an example application of the implemented function to predict the Higgs boson
pair production cross-section in the 
\GW{Two-Higgs-}Doublet model~(2HDM) \GW{of} type~II.
The 2HDM predicts a second \cp-even
Higgs boson 
$H$ that can be resonantly produced
and, if the mass $m_H$ is sufficienty large,
decay into a 
pair of SM-like Higgs bosons $h$.
We 
use a parameter point \GW{where} the 2HDM parameters
(see, e.g.~\ccite{Biekotter:2023eil} for the definitions
of the model parameters)
\GW{are} fixed to $\tan\beta = 2$, $\cos(\beta-\alpha)=0.1$, $m_H=m_A=m_{H^{\pm}} = M \equiv \sqrt{\sin\beta\,\cos\beta\, m_{12}^2}$ 
and plot the prediction of \HiPr for a variable resonant mass $m_H$. We use the public tool \texttt{thdmTools}~\cite{Biekotter:2023eil} (interfaced to
\texttt{HDECAY}~\cite{Djouadi:1997yw,Djouadi:2018xqq})
to compute the total width $\Gamma_H$ of the 
boson $H$. We use the following tree level formulas to compute the
relevant coupling coefficients,
\begin{align}
\label{eq:coups_thdm}
    \GW{c_{t}} &= \sin(\beta-\alpha) + \cos(\beta-\alpha) \cot \beta,\\
    c_{t}^{H} &= \cos(\beta-\alpha) - \sin(\beta-\alpha) \cot \beta,\\
  \lambda_{hhh}
        &= \frac{1}{2v^2} \bigg\{ m_h^2 s_{\beta -\alpha }^3 
        +\left(3 m_h^2-2 M^2\right) c_{\beta -\alpha }^2 s_{\beta -\alpha }
        +2 \cot 2 \beta \left( m_h^2-M^2\right) c_{\beta -\alpha
        }^3\bigg\}, \\[.5em]
    \lambda_{hhH}
        &= -\frac{\cos(\beta-\alpha)}{2v^2} \bigg\{  \left(2m_h^2+m_H^2-4M\right)\sin(\beta-\alpha)^2
        -\left(2m_h^2+m_H^2-2M \right) \cos(\beta-\alpha)^2 \nonumber\\
        &+2\cot{2\beta}\left(2m_h^2+m_H^2-3\TB{M}\right) \sin(\beta-\alpha)\cos(\beta-\alpha)
         \bigg\} \, .
\end{align}
Here the coefficients are normalised
such that the \GW{input} couplings for the \HiPr function
\texttt{HP.EffectiveCouplingCxns.ggHH}
are $c_t$, $c_3 = c_{t}^{H} \times \lambda_{hhH}$ and $\kappa_{\lambda}=\lambda_{hhh}/\lambda_{hhh}^{\mathrm{SM},(0)}$, where
\GW{$\lambda_{hhh}^{\SH{\mathrm{SM},(0)}}$ refers to the lowest-order coupling in the SM
(see \cref{eq:kaladef}),}
$\lambda_{hhh}^{\mathrm{SM},(0)} = m_h^2/(2\,v^2)\approx 0.13$.

The total 
Higgs boson pair production cross-section prediction is shown with the 
solid black line in \reffi{fig:gghh_thdm}, \GW{where} no experimental constraints are applied. 
The non-resonant contribution is indicated in light green, while the resonant-only contribution is shown in dark green. A significant enhancement of the cross-section is observed for values of the heavy scalar mass $m_H$ near the di-top mass threshold, where the resonant contribution overwhelmingly dominates the total production cross-section. For BSM masses above approximately 700 GeV, the non-resonant contribution becomes the dominant one, causing the total cross-section to \GW{reach a} plateau at the value predicted by the SM. In this 
\GW{example scenario where all BSM masses are degenerate}, the interference effects (as seen by the difference between the dark green and the black lines) are small. \GW{However,} this is not true in general. \GW{The non-resonant and 
interference contributions} in di-Higgs production can play a \GW{very} significant role \GW{regarding the limits that can be obtained from the resonant di-Higgs searches~\cite{Heinemeyer:2024hxa}.
This is in particular the case if a non-vanishing splitting between the BSM mass scales induces a large deviation in $\kappa_\lambda$ from the SM prediction~\cite{Bahl:2022jnx,Bahl:2023eau,Heinemeyer:2024hxa}.}

We also display the LO prediction from \texttt{anyHH} in orange, alongside the NLO prediction from \texttt{HPAIR} in the heavy top limit \GW{(blue)}.
Both codes include the resonant and non-resonant
contributions and the interference effects between those.
A discrepancy of roughly a factor of two is observed between these two predictions,
due to the well-known 
NLO QCD $K$-factor of about $K \sim 2$~\cite{Djouadi:1991tka}.
The inclusion of the NNLO QCD 
corrections in the resonant 
gluon fusion production
cross-section of the BSM state~$H$, as detailed above, further raises the $K$-factor w.r.t.\ the 
LO prediction to approximately a factor of $K \sim 3$
in the range of $m_H$ where the resonant production
is relevant. 
The QCD NNLO corrections also explain
the enhancement compared
to the prediction from \texttt{HPAIR} that
is visible for $m_H \lesssim 500\gev$.



\section{\HiBo}
\label{sec:hb}


In this \GW{section}, we discuss updates to \HiBo. The main improvements are the inclusion of searches
using 
multi-top final states with up to four top quarks, 
coupling-dependent non-resonant di-Higgs limits, and \GW{an improved} handling of pre-Higgs discovery searches.


\subsection{BSM Higgs searches with a multi-top final state}
\label{sec:multi_top}

Since the coupling of Higgs bosons to fermions
are proportional to the respective fermion mass,
BSM scalars with large top-quark Yukawa couplings appear in 
\GW{many}
extensions of the SM Higgs sector. This makes so-called multi-top final states a natural 
candidate for searches for these scalars
if their masses are larger than twice the top-quark mass.
At the LHC, the most relevant production channels for a scalar $H$ with a large top-quark Yukawa coupling are 
production via gluon fusion, $ggH$, associated
production with a top quark, $tH$, and with
a top-quark and a $W$-boson, $tWH$, 
as well as associated production with
a top-quark pair, $t\bar t H$. 
Assuming the scalar to decay into two top quarks, the corresponding final states consist of two, three, or four top quarks. 
While di-top searches utilizing first-year Run~2
data~\cite{CMS:2019pzc} 
have already been implemented in \HiBo, no searches for final states with three or four top quarks were
implemented yet.\footnote{\TB{The more recent CMS~\cite{CMS:2025dzq} and ATLAS~\cite{ATLAS:2024vxm} searches in the di-top final state, based on the
full Run~2 data, resulted in coupling limits
on BSM resonances that are in strong contradiction
with each other,
in particular for BSM resonances
with masses close to the di-top threshold.
These limits have therefore not yet been implemented
in \HiBo.}}
Here, we describe the recasting of the CMS 
search for the production of four top 
quarks~\cite{CMS:2019rvj} and its implementation in \HiBo. We leave the implementation of the corresponding ATLAS search~\cite{ATLAS:2024jja}, which gave rise to
comparable cross-section limits, for future work.

\paragraph{Effective model description}

For our recasting/implementation into \HiBo, we use a simplified model approach in close analogy to the Higgs-characterization model defined in \ccite{Artoisenet:2013puc}. The Yukawa part of the Lagrangian for the top quark
is given by
\begin{equation}
\mathcal{L}_\text{Yuk} = -\frac{y_t^\text{SM}}{\sqrt{2}}\bar{t}(c_t + i \gamma_5 \tilde{c}_t)t \phi.
\end{equation}
Here, $y_t^{\mathrm{SM}}$ is the SM top-Yuakwa coupling, $\phi$ denotes a generic scalar with mass $m_\phi$, $t$ denotes the top-quark field,
and $c_t$ and $\tilde{c}_t$ are the \cp-even and 
\cp-odd top-quark Yukawa coupling modifiers, respectively. For the scalar $\phi$, we also consider the vector boson couplings
\begin{equation}
\mathcal{L}_{V} = c_V \phi \left(\frac{M^2_Z}{v}Z_\mu Z^\mu + 2\frac{M^2_W}{v}W^+_\mu W^{-\mu}\right),
\end{equation}
where $Z$ and $W$ denote the massive vector-boson fields with the masses $M_{Z}$ and $M_W$, respectively.
The coupling coefficient
$c_V$ rescales the vector-boson coupling of $\phi$ with respect to the
couplings of the Higgs boson predicted by the SM.


\paragraph{Cross-section comparison of \texorpdfstring{\boldmath{$t\bar t\phi$}}{ttphi}, \texorpdfstring{$tW\phi$}{tWphi} and \texorpdfstring{$t\phi$}{tphi} at the LHC}

\begin{figure}[!htb]
	\centering
 \includegraphics[width=\textwidth]{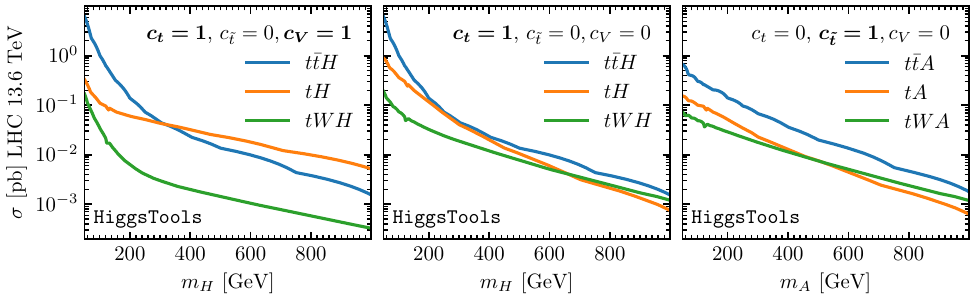}
 \caption{Comparison of the total $t\bar tH$, $tWH$ and $tH$ cross-sections
 for three different coupling configurations.}
 \label{fig:4t_xsec_comp}
\end{figure}

Before we discuss the recasting of \ccite{CMS:2019rvj}, we compare in \reffi{fig:4t_xsec_comp} the production cross-sections of the 
discussed $t\bar t\phi$, $tW\phi$ and $t\phi$ production modes at the LHC for a heavy scalar as a function of the mass. 
The left panel shows the production cross-sections for a SM-like Higgs boson; the middle panel, for a \cp-even BSM scalar with $c_t = 1$ and $c_V = 0$ (as it can arise,
for example, in the alignment limit of the 2HDM); and, the right panel, for a \cp-odd BSM scalar with $c_{\tilde{t}} = 1$. While the $t\bar t\phi$ cross-section (blue lines) dominates over the $t\phi$ (orange lines) and $tW\phi$ (green lines) cross-sections for a SM-like Higgs boson with a mass below $\sim 350\gev$ (due to a large 
\GW{destructive}
interference between the contribution proportional to the top-quark Yukawa coupling and the contribution proportional to the $H W^+ W^-$ \GW{vertex}), 
\GW{the relative rates for the three cross-sections are in general different for BSM Higgs bosons}. For masses above $\sim 350\gev$, the $t\phi$ cross-section is larger than the $t\bar t \phi$ cross-section for a SM-like Higgs boson. 
\GW{The comparison with the two other coupling configurations shows that}
for scalars with \GW{$c_V \approx 0$}, the $t\phi$ and $tW\phi$ cross-sections give substantial contributions to top-associated Higgs \TB{boson}
production throughout the considered mass range.
This shows that for BSM searches, $t\phi$ and $tW\phi$ production can give sizeable contributions and should not be neglected (as done e.g.\ in \ccite{ATLAS:2024jja}).



\subsubsection{Recasting}
\label{sec:recast}

In the following, we will discuss the recasting of \ccite{CMS:2019rvj}.
\TB{Using} the \texttt{MadAnalysis}~\cite{Conte:2014zja,Dumont:2014tja,Araz:2019otb} implementation of \ccite{CMS:2019rvj} presented in \ccite{Darme:2021gtt}, we fit the number of expected signal events in the most sensitive signal region as a function of the BSM scalar mass and the coupling modifiers $c_t$, $\tilde c_t$, and $c_V$.
Our implementation is subject to several restrictions. CMS uses a boosted-decision-tree analysis to optimize their analysis. We are limited to using their supplementary cut-based analysis which divides the events in fourteen signal regions\TB{, where w}e treat the total number of events, which is the SM background plus SM $t\bar{t}t\bar{t}$ events, as background for our implementation. Furthermore,
the correlations between the signal regions
\TB{are not publicly available, such that we} 
are restricted to derive our limits from the events in the most sensitive signal region, which we find to be signal region 8,
\GW{as}
\TB{defined in \ccite{CMS:2019rvj}},
for all considered parameter points.
We \TB{also} do not account for signal-background interference
effects, which is justified in searches targeting final
states with more than two top quarks, where interference with 
QCD background processes is significantly less relevant than
in di-top final states.
Moreover, in scenarios with multiple BSM scalars decaying into
top quarks, 
\TB{we sum their} contributions 
incoherently,
without accounting for signal–signal interference,
as is discussed \TB{in more detail} below.
In principle, the signal yields also depend on the decay width 
\TB{$\Gamma_\phi$} of the BSM scalar $\phi$, which we 
\GW{keep} as
\TB{an independent parameter in our}
implementation into \HiBo.
\TB{However, we neglect the width-dependence of the
signal yields.}
To justify this simplification, we computed the signal yields for different width values of the decaying Higgs boson ($1\%$, $5\%$, $10\%$ and $15\%$ of $m_\phi$), finding only negligible differences of a few percent 
\GW{in the signal efficiencies,}
demonstrating that
\GW{the derived limits on cross-section times branching ratio are}
effectively independent of the decay width.

\paragraph{Signal yield fit formulas}

While the search of \ccite{CMS:2019rvj} mainly targets
final states with four top quarks, originating from $t\bar t \phi$ production, also three top-final states originating from $t\phi$ or $tW\phi$ production can have a non-negligible contribution \GW{as discussed above}. In contrast, we found the contribution from di-top final states to be negligible and therefore do not consider it in the recasting.
\GW{This leads,}
\TB{for a fixed value of $m_\phi$,}
\GW{to a parametrisation of}
the total cross-section for each process $X \TB{ = \{ t \bar t \phi, t W \phi, t \phi \} }$ in the form
\begin{align}
\sigma_{\text{tot},X} &= (a_{1,\text{tot},X} c_V^2 + a_{2,\text{tot},X} c_V c_t + a_{3,\text{tot},X} c_t^2 + a_{4,\text{tot},X} \tilde c_t^2)\cdot(b_{1,\text{tot},X} c_t^2 + b_{1,\text{tot},X} \tilde c_t^2) = \nonumber\\
&= c_{1,\text{tot},X} c_V^2 c_t^2 + c_{2,\text{tot},X} c_V^2 \tilde c_t^2 + c_{3,\text{tot},X} c_V c_t^3 + c_{4,\text{tot},X} c_V c_t \tilde c_t^2 + c_{5,\text{tot},X} c_t^4 \nonumber\\
&\hspace{.35cm}+ c_{6,\text{tot},X} c_t^2 \tilde c_t^2 + c_{7,\text{tot},X} \tilde c_t^4\,,
\end{align}
where the first bracket comes from the production and the second bracket from the decay.
All other possible coefficients are zero as a result of the non-interference between \cp-even and \cp-odd contributions. 
In the same way, we can also parametrise the \GW{signal yield}
in the most sensitive signal region, which is obtained by multiplying the total cross-section by the acceptance and efficiency, in the form
\begin{align}
\sigma_{\epsilon,X} &= c_{1,\epsilon,X} c_V^2 c_t^2 + c_{2,\epsilon,X} c_V^2 \tilde c_t^2 + c_{3,\epsilon,X} c_V c_t^3 + c_{4,\epsilon,X} c_V c_t \tilde c_t^2 + c_{5,\epsilon,X} c_t^4 + c_{6,\epsilon,X} c_t^2 \tilde c_t^2 + c_{7,\epsilon,X} \tilde c_t^4
\end{align}
for each process $X$.

\TB{In dedicated Monte-Carlo simulations
it was} found that in the dominant 
contribution with four \TB{final state} top quarks, the presence of more than one scalar \GW{particle} leads to constructive
signal--signal interference~\cite{Bahl:2026ama}.
To simplify our implementation, we make the choice to neglect 
\GW{such} interference terms. Simply adding the yields from 
\GW{new scalar particles} slightly underestimates the signal efficiency and leads to a conservative limit setting. We then use \texttt{MadGraph} and the \texttt{MadAnalysis} implementation of \ccite{Darme:2021gtt} to fit the coefficients $c_{1-7,\TB{\epsilon,X}}\TB{(m_\phi)}$ in each of the signal regions, for each contributing process
$X$, \TB{and for different mass values
in the range
$350\gev \leq m_\phi \leq 1\tev$}.\footnote{\TB{New resonances
with masses at the multi-TeV scale can be more efficiently
searched for in final states with boosted
top quarks~\cite{Darme:2025leu}.}}


\begin{figure}[!htb]\centering
\includegraphics[width=.48\textwidth]{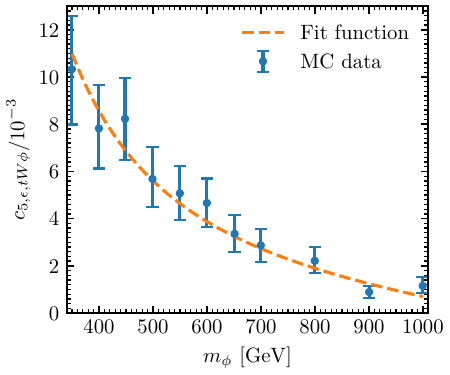}\hfill 
\includegraphics[width=.48\textwidth]{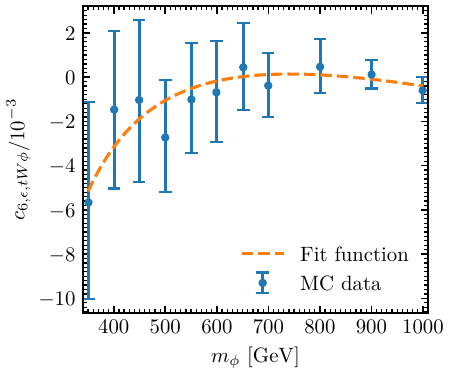}
\caption{Exemplary fit functions (orange dashed) of the $tW\phi$ coupling coefficients $c_5$ (left) and $c_6$ (right) plotted 
\GW{in comparison to}
the original Monte-Carlo data
for signal region 8.}
\label{fig:examplefit}
\end{figure}

In \reffi{fig:examplefit}, we show two exemplary fit functions (orange dashed) and the underlying \GW{Monte-Carlo} data points (blue) 
\GW{where}
the error bars 
\GW{indicate} statistical uncertainties
from the limited number of Monte-Carlo events.
In the left panel, we show the fit function for the coefficient 
\GW{$c_{5,\epsilon,tW\phi}$},
which corresponds to a vertex proportional to  $c_t^4$ in signal region~8.
The function, whose form is chosen heuristically, is given by
\TB{
\begin{equation}
c_{5,\epsilon,tW\phi}(m_\phi) =
   \frac{1.2 \cdot 10^{3}}{m_\phi^2} - 2.6 \cdot 10^{-5} \ m_\phi + 2.1 \cdot 10^{-3} \ .
\end{equation}
As a second example,
in the right panel
the results for the coefficient
$c_{6,\epsilon,tW\phi}$ \GW{are shown},
which correspond to the term proportional
to $c_t^2 \tilde{c}_t^2$.}
\GW{The (heuristically chosen)}
fit functions \TB{for all
three production processes}
are listed in \cref{app:tttt_fit_functions}.

\begin{figure}[!htb]\centering
\includegraphics[width=.48\textwidth]{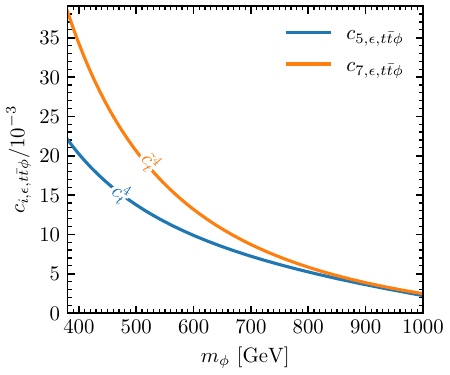}\hfill 
\includegraphics[width=.48\textwidth]{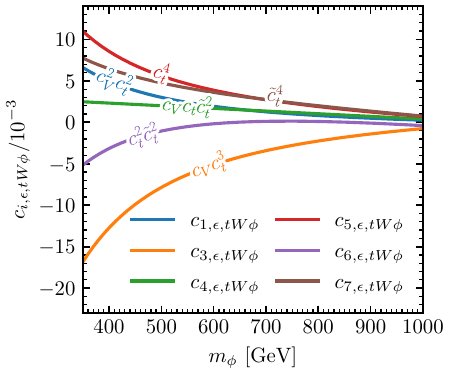}
\caption{Contributing fit functions for $t\bar t \phi$ (left) and $tW\phi$ (right) \GW{production}.
}
\label{fig:fitgather}
\end{figure}

To \TB{compare} 
the contributions of each coefficient to the
cross-section calculation, we \TB{show} 
\TB{their fit functions}
in \reffi{fig:fitgather} for $t\bar t \phi$ (left panel) and $tW\phi$ (right panel) \GW{production}. The $t\bar t \phi$ channel only has contributions from $c_{5,\epsilon,t \bar t \phi}$ and $c_{7,\epsilon,t \bar t \phi}$, while \TB{we found the fit function for} $c_{6,\epsilon,t \bar t \phi}$
\TB{to be compatible with zero across the whole
mass range within the statistical uncertainties,}
and $c_{1-4,\epsilon,t \bar t \phi}$ are zero because there is no vertex containing $c_V$ in $t\bar t \phi$ production. For the $tW\phi$ channel, 
\GW{there are significant negative} contributions from
\TB{$c_{3,\epsilon,tW\phi},c_{6,\epsilon,tW\phi} < 0$}
and positive contributions from
\TB{$c_{1,\epsilon,tW\phi},c_{4,\epsilon,tW\phi},
c_{5,\epsilon,tW\phi},c_{7,\epsilon,tW\phi} > 0$,}
while only $c_{2,\epsilon,tW\phi}$ is equal to zero. This indicates relevant contributions from vertices which are proportional to $c_V$ in the coefficients
\TB{$c_{1,\epsilon,tW\phi},c_{3,\epsilon,tW\phi},
c_{4,\epsilon,tW\phi}$.}
Although some of the coefficients are negative, we checked that the total signal yields stay positive
\TB{across the considered mass range.}

\subsubsection{Validation: cross-section and \texorpdfstring{tan\pdfmath{\beta}}{tanb} limits}
\label{sec:validtanbeta}

The \HiBo limit is implemented in the form of an upper limit on the number of events in the most sensitive signal region, which we found to be signal region~8
\TB{used in the CMS analysis~\cite{CMS:2019rvj}}. 
For validation, we compare to the 
\GW{upper} limit on $\sigma(t\bar t\phi + t\phi + tW\phi)\cdot\text{BR}(\phi\to t\bar t)$ \GW{obtained by CMS}. 
As an example, to derive a limit on the production rate for a scalar with $\tilde c_t = c_V = 0$, we can rewrite the production rate as (keeping the total width again as an independent parameter)
\begin{equation}
\sigma(t\bar t\phi + t\phi + tW\phi)\cdot\text{BR}(\phi\to t\bar t) = c_t^4 \frac{\left[\sigma(t\bar t\phi + t\phi + tW\phi)\cdot \Gamma(\phi\to t\bar t)\right]_{c_t=1}}{\Gamma_\text{tot}}
\end{equation}
given the scaling with $c_t^4$.
The number of signal events is then given by
\begin{equation}
N_\text{signal} = c_t^4 \ \mathcal{L} \cdot \left[\sigma(t\bar t\phi,\phi\to t\bar t)\epsilon_{t\bar t \phi} + \sigma(t\phi,\phi\to t\bar t)\epsilon_{t \phi} + \sigma(tW\phi,\phi\to t\bar t)\epsilon_{tW \phi}\right]_{c_t=1}\,,
\end{equation}
where $\mathcal{L}$ is the luminosity and $\epsilon_X \equiv \sigma_{\epsilon,X}/\sigma_{\text{tot},X}$
are the efficiencies for the channel $X$\TB{, and
the cross-sections $\sigma_{\epsilon,X}$
and $\sigma_{\text{tot},X}$ are determined
as discussed above}.
We obtain a limit on the number of signal events from the observed number of events in the considered signal region using \texttt{MadAnalysis}.
This allows us to test the couplings and masses of a given model point against the limit on the number of signal events and derive a limit on the couplings and \GW{the} cross-section. Similarly, we 
\GW{obtain}
the number of signal events for all coupling configurations with $\tilde c_t \neq 0$ and $c_V \neq 0$.


\begin{figure}
\centering
\includegraphics[width=0.48\textwidth]{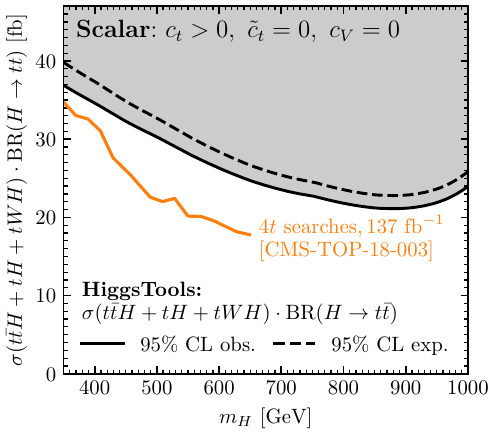}~
\includegraphics[width=0.48\textwidth]{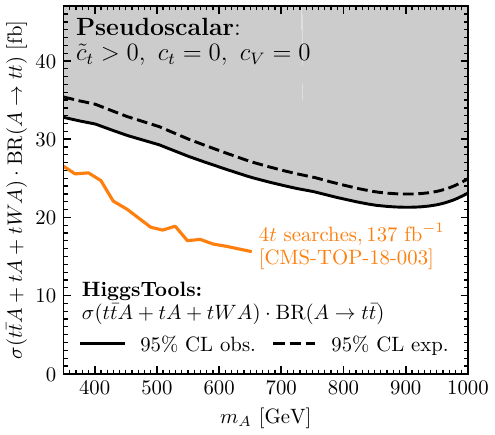}
\caption{
\GW{Expected} (dashed) and observed (solid) 95\% CL limits on the total cross-sections obtained from the recasting of the CMS searches in the $t \bar t t \bar t$ final state implemented in \HiBo in black,
and the corresponding observed limit reported by CMS~\cite{CMS:2019rvj} in orange.
The results for a \cp-even scalar that only couples to top quarks ($c_V = 0$) are shown in the left plot, and the results for a corresponding \cp-odd pseudoscalar are shown in the right plot.}
\label{fig:xsec_validation}
\end{figure}

In \cref{fig:xsec_validation} we show the cross-section limit for a 
\GW{\cp-even scalar with $c_V = 0$}
(left panel) and a pseudoscalar (right panel) obtained from \HiBo in comparison to the limits from CMS (orange). As expected, because of our 
\GW{restriction}
to the most sensitive signal region,
\TB{and since we are limited to the cut-based analysis,}
our limit is overall 
\GW{somewhat}
weaker compared to the CMS result for the total cross-section times branching ratio $\sigma(pp \to (tt,tW,t)+\phi) \cdot \mathrm{BR}(\phi \to t\bar{t})$.
\TB{However, due to our recasting in which we
consider a broader mass range and a more general
coupling structure, our limit} 
\GW{is applicable up}
to higher masses,
\TB{and it} 
\GW{can be used for}
\cp-mixed scalars and allows for a non-zero
coupling to vector bosons. 


\begin{figure}[!htpb]
\centering
\includegraphics[width=0.328\textwidth]{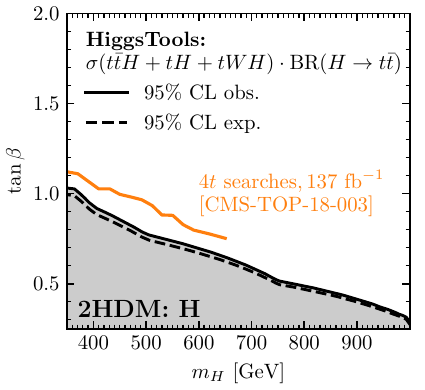}
\includegraphics[width=0.328\textwidth]{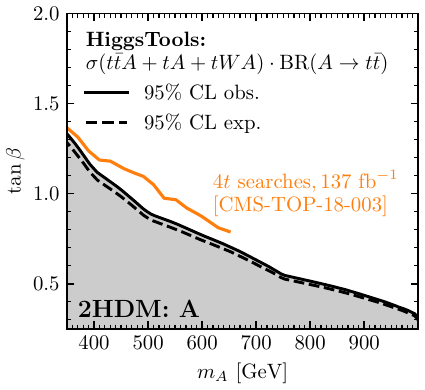}
\includegraphics[width=0.328\textwidth]{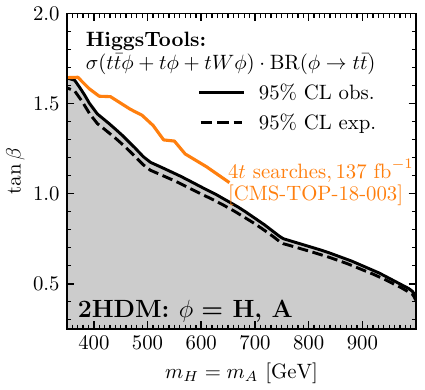}
\caption{Observed (black \GW{solid}) and expected (black dashed) lower limits on tan$\beta$ compared to the observed limit from the original CMS analysis (orange) for a 2HDM scalar~$H$ (left), pseudoscalar~$A$ (centre) and scalar + pseudoscalar of the same mass (right).} 
\label{fig:tanblimval}
\end{figure}

In \cref{fig:tanblimval}, we compare the upper limits on $\tan\beta$ in the type-I 2HDM obtained from our \HiBo implementation to the 
\GW{limits obtained by CMS}.
The left panel shows the case of a single scalar, the middle plot \GW{the case} of a single pseudoscalar, and the right panel the case of a scalar and a pseudoscalar with the same mass. The grey area indicates the excluded 
\GW{region}. The dashed black lines indicate the corresponding expected limit. The orange lines show the observed limit from the original CMS analysis. Overall, we 
\GW{find}
good agreement with the CMS results given the restrictions outlined above.




\subsubsection{Example \GW{applications}}

In this subsection, we discuss several example applications of the newly implemented four-top limit.


\paragraph{The two Higgs doublet model}
\label{sec:thdm}

\begin{figure} 
\centering
\includegraphics[width=0.49\textwidth]{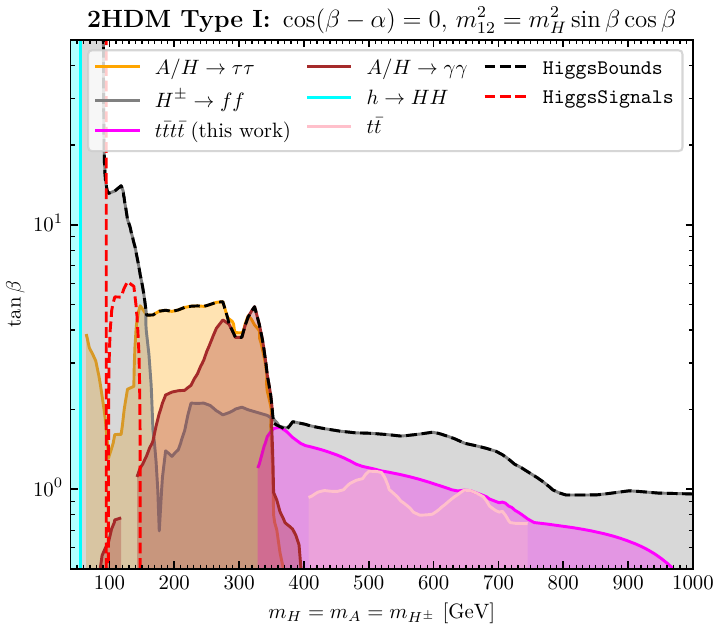}~
\includegraphics[width=0.49\textwidth]{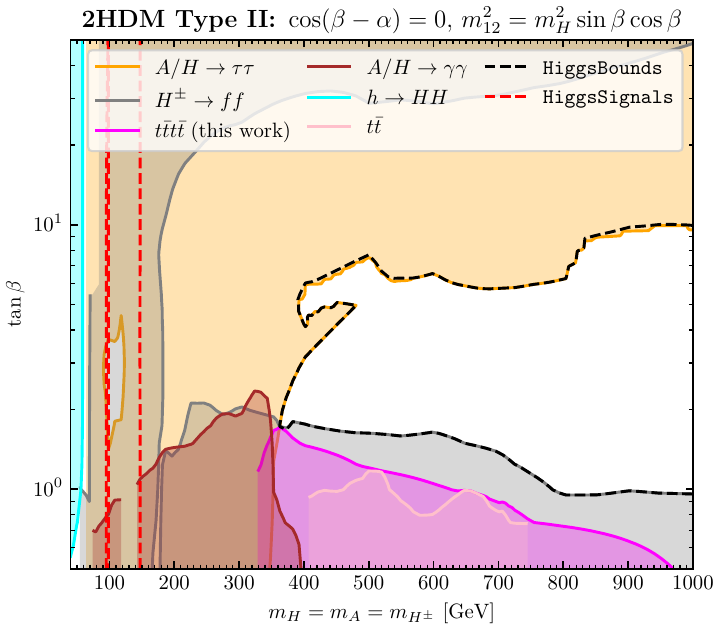}\\[0.4em]
\includegraphics[width=0.49\textwidth]{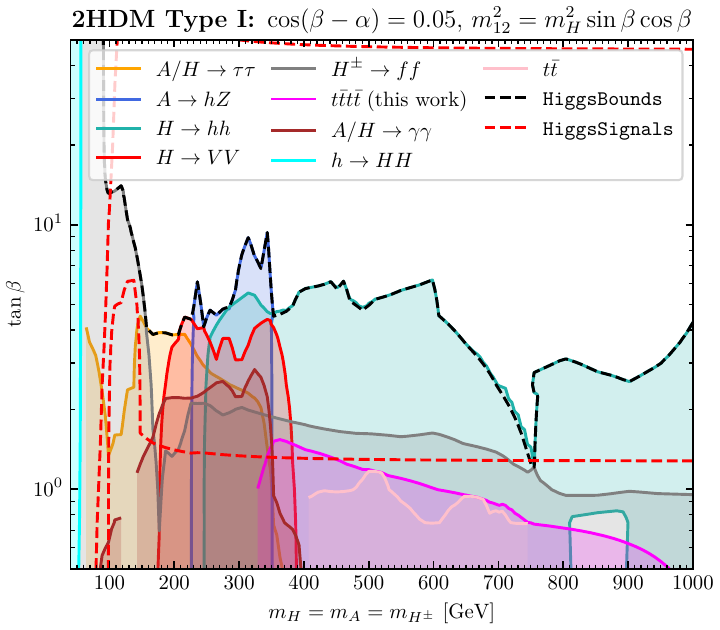}~
\includegraphics[width=0.49\textwidth]{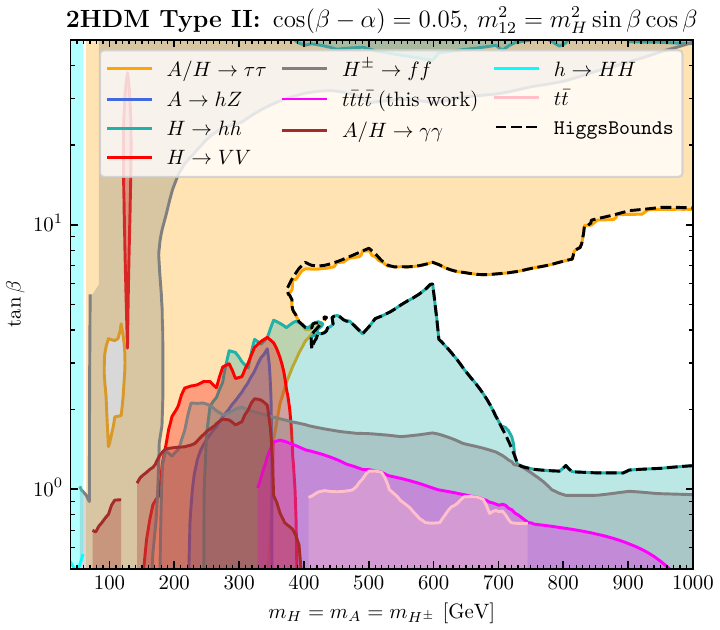}\\[0.4em]
\includegraphics[width=0.49\textwidth]{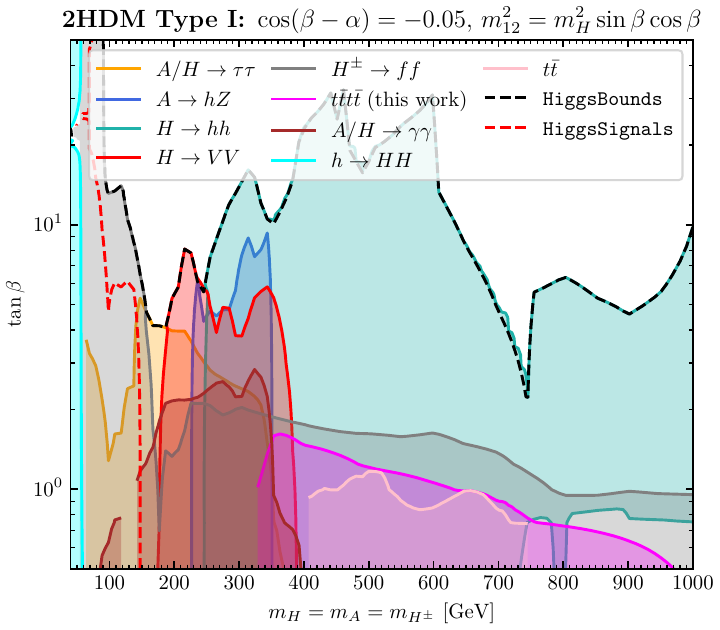}~
\includegraphics[width=0.49\textwidth]{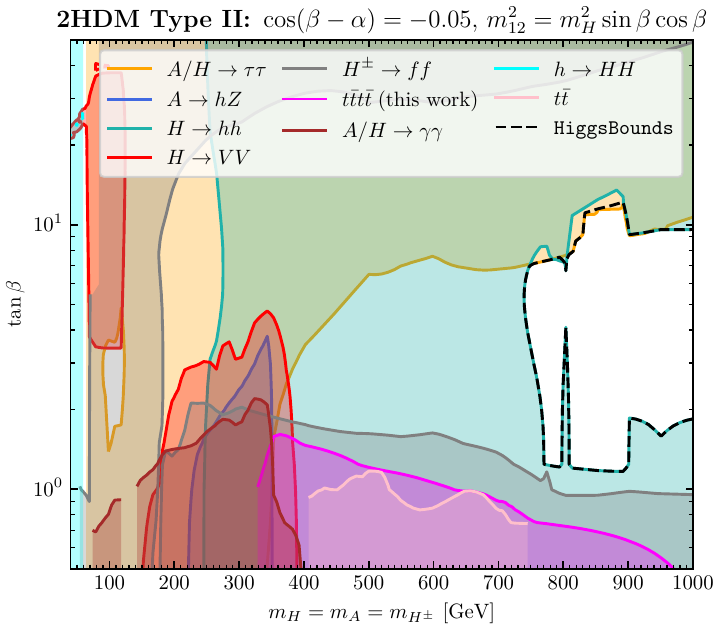}
\caption{95\% CL exclusion regions from
all searches implemented in \HiTo
in 2HDM parameter planes with
the degenerate BSM scalar masses on
the horizontal axis and $\tan\beta$
on the vertical axis. The left and right plots
show the exclusion regions in the
\GW{2HDM of}
type~I and~II, respectively. The remaining 2HDM parameters
are set as shown on top of
each plot. Different coloured regions
show the exclusion regions from individual
searches in 
the final states specified in the legends.
The dashed black \KR{(red)} line indicates the
combined \HiTo~\KR{(\HiSi)} exclusion region.}
\label{fig:thdmplanes}
\end{figure}

Here, we discuss the impact of these new limits on the 2HDM parameter space, and we compare the corresponding exclusion regions to the ones from \TB{other} searches for neutral and charged Higgs bosons performed at LEP and the LHC. In \cref{fig:thdmplanes}, we show six plots with coloured exclusion regions from various searches in a parameter plane defined by the degenerate BSM scalar masses $m_H = m_A = m_{H^\pm}$ on the horizontal axis and the parameter $\tan\beta$ on the vertical axis. \TB{The \cp-even
Higgs boson $h$ 
\GW{corresponds to}
the SM-like Higgs
boson $h$} \GW{in the considered scenarios}. The three plots in the left column are for the 2HDM of type~I, and the three plots in the right column are for the 2HDM of type~II. 
\GW{The} plots in the top, middle and bottom rows correspond to different values of $\cos (\beta - \alpha) = 0.0, 0.05$ and $-0.05$, respectively, and the remaining free parameter $m_{12}^2$ is fixed via the relation $m_{12}^2 = m_H^2 \sin\beta \cos\beta$ for all shown parameter planes.


One can see in the top row for $\cos (\beta - \alpha) = 0$ --- i.e., the alignment limit in which the Higgs boson $h$ at 125~GeV resembles a SM Higgs boson \TB{if no BSM decay mode is kinematically allowed} --- that the newly implemented cross-section limits from \GW{the} $\tttt$ searches give rise to exclusion regions at low $\tan\beta$ values (magenta) that are comparable to the exclusions based on searches for neutral scalars in the di-top final state (pink) and slightly weaker than searches for charged scalars decaying into a top-quark and a bottom-quark (grey). In the plots in the middle row, one can see that for a positive value of $\cos (\beta - \alpha) = 0.05$ the more powerful exclusion regions arise from searches for resonant Higgs-boson pair production via $H \to hh$ decays, which can exclude significantly larger $\tan\beta$ values \TB{than the $t \bar t t \bar t$ searches}. The same observation can be made in the plots in the bottom row for a negative value of $\cos(\beta - \alpha) = -0.05$. 
\KR{The spike at around $800 \gev$ is an artifact of the selection of a different search in that region because of a larger expected ratio. In particular, the resonant di-Higgs ATLAS combination \cite{ATLAS:2023vdy} is selected, while to the sides of the spike the resonant di-Higgs search in the $b\bar{b}\gamma\gamma$ final state \cite{ATLAS:2022xzm} is selected as the most sensitive one, causing a visible discontinuity in the allowed region.}
We summarize the LHC searches contributing to the depicted exclusion regions in \cref{tab:thdmsearches}.

\begin{table}[t]
  \centering
  {\scriptsize
  \renewcommand{\arraystretch}{1.2}
  \begin{tabular}{l||l|l|l|l}
    \textbf{Channel} & \textbf{Final states}
      & \textbf{Energy}
      & \textbf{Int.~luminosity}
      & \textbf{References} \\
    \hline
    \hline
    $A/H \to \tau^+ \tau^-$ &
      &
      7 TeV + 8 TeV &
      4.9~fb$^{-1}$ + 24.6~fb$^{-1}$ &
      CMS: \cite{CMS-PAS-HIG-14-029} \\
    &
      &
      13 TeV &
      139~fb$^{-1}$ &
      ATLAS: \cite{ATLAS:2020zms},
        CMS: \cite{CMS:2022goy} \\
    &
      &
      13 TeV &
      35.9~fb$^{-1}$ &
      CMS: \cite{CMS:2018rmh} \\
    \hline
    $H^\pm \to f f$ &
      $tb$ &
      13 TeV &
      139~fb$^{-1}$ &
      ATLAS: \cite{ATLAS:2021upq} \\
    &
      $tb$ &
      13 TeV &
      35.9~fb$^{-1}$ &
      CMS: \cite{CMS:2020imj} \\
    &
      $\tau \nu$ &
      13 TeV &
      36~fb$^{-1}$ &
      ATLAS: \cite{ATLAS:2018gfm}
        CMS: \cite{CMS:2019bfg} \\
    &
      $\tau \nu$, $cs$ &
      189--209 GeV &
      2.6~fb$^{-1}$ &
      LEP: \cite{ALEPH:2013htx} \\
    \hline
    \rowcolor{magenta!30!}
    $pp \to \tttt$ &
      &
      13 TeV &
      137~fb$^{-1}$ &
      recasting of CMS: \cite{CMS:2019rvj} \\
    \hline
    $A/H \to \gamma\gamma$ &
      &
      8 TeV &
      20.3~fb$^{-1}$ &
      ATLAS: \cite{ATLAS:2014jdv} \\
    &
      &
      13 TeV &
      139~fb$^{-1}$ &
      ATLAS: \cite{ATLAS:2021uiz} \\
    \hline
    $H \to hh$ &
      $b \bar b b \bar b$, $b \bar b \tau^+ \tau^-$,
        $b \bar b \gamma \gamma$ &
      13 TeV &
      139~fb$^{-1}$ &
      ATLAS: \cite{ATLAS:2023vdy} \\
    &
      $b \bar b \tau^+ \tau^-$ &
      13 TeV &
      139~fb$^{-1}$ &
      ATLAS: \cite{ATLAS:2022xzm} \\
    &
      $b \bar b \gamma \gamma$, $b \bar b \tau^+ \tau^-$,
        $b \bar b b \bar b$, $b \bar b VV$ &
      13 TeV &
      35.9~fb$^{-1}$ &
      CMS: \cite{CMS:2018ipl} \\
    &
      $b \bar b b \bar b$ (boosted) &
      13 TeV &
      139~fb$^{-1}$ &
      CMS: \cite{CMS:2024pjq} \\
    &
      $b \bar b \gamma \gamma$, $b \bar b \tau^+ \tau^-$,
        $b \bar b W^+ W^-$, &
      13 TeV &
      138~fb$^{-1}$ &
      CMS: \cite{CMS:2024phk} \\ 
    &
      $W^+ W^- W^+ W^-$, $\tau^+ \tau^- \tau^+ \tau^-$, &
      &
      &
      \\
    &
      $W^+ W^- \tau^+ \tau^-$, $W^+ W^- b \bar b$ &
      &
      &
      \\
    \hline
    $A/H \to t \bar t$ &
      &
      13 TeV &
      35.9~fb$^{-1}$ &
      CMS: \cite{CMS:2019pzc} \\
    \hline
    $A \to ZH$ &
      $\nu \bar \nu b \bar b$, $\ell^+ \ell^- b \bar b$ &
      13 TeV &
      139~fb$^{-1}$ &
      ATLAS: \cite{ATLAS:2022enb} \\
    &
      $\nu \bar \nu b \bar b$, $\ell^+ \ell^- b \bar b$ &
      13 TeV &
      35.9~fb$^{-1}$ &
      CMS: \cite{CMS:2019qcx} \\
    \hline
      $H \to VV$ &
        $\ell^+ \ell^- \ell^{'+} \ell^{'-}$,
          $\ell^+ \ell^- \nu \bar \nu$ &
      13 TeV &
      139~fb$^{-1}$ &
      ATLAS: \cite{ATLAS:2020tlo} \\
    &
      $\ell^+ \ell^- \ell^{'+} \ell^{'-}$,
        $\ell^+ \ell^- \nu \bar \nu$,
        $\ell^+ \ell^- q \bar q$ &
      13 TeV &
      35.9~fb$^{-1}$ &
      CMS: \cite{CMS:2018amk} \\
    \hline
    $h \to HH/AA$ &
      $b \bar b \tau^+ \tau^-$ &
      13 TeV &
      35.9~fb$^{-1}$ &
      CMS: \cite{CMS:2018zvv} \\
  \end{tabular}
  \renewcommand{\arraystretch}{1.0}
  }
  \caption{Experimental searches for new Higgs bosons
  whose cross-section limits give rise to the
  exclusion regions shown in \cref{fig:thdmplanes}.
  The row highlighted in magenta corresponds to
  the CMS measurements of the production of
  four top quarks that is recasted in this work
  in order to set limits on the production
  of heavy neutral Higgs bosons.}
  \label{tab:thdmsearches}
\end{table}

In addition to the exclusion regions from LHC searches for additional Higgs bosons, we indicate in \cref{fig:thdmplanes} with the red dashed lines the exclusion regions resulting from \HiSi. 
\TB{According to our discussion in \cref{sec:chisqrecom} \GW{below},}
we apply as a condition for valid parameter regions $\Delta\chi^2_{125} \leq 6.18$, which corresponds to a joint parameter estimation with two free parameters at $2\sigma$ confidence level assuming a Gaussian $\chi^2$-distribution.\footnote{We checked that in this way our exclusion regions are in good agreement with the exclusion regions obtained by ATLAS and CMS from global fits taken into account their respective Run~1 and Run~2 datasets.} In each plot, the area below and to the left of the red dashed line is excluded, with the exception of the left plot in the middle row, where at $\tan\beta \approx 40$ an additional red dashed line is visible which indicates an additional exclusion region above this line. In the middle and bottom plots in the right column for the Yukawa type~II, no red dashed line is visible because 
the whole parameter planes 
\GW{are} excluded by 
\GW{the \HiSi\ constraints based on the current}
measurements of the 125~GeV Higgs boson.


\paragraph{The top-philic ALP}
\label{sec:alp}

As a second example, we apply the limits from our recasting of the CMS search in the $\tttt$ final state to the case of an axion-like particle~(ALP) that couples to top quarks. In the non-derivative basis, the relevant ALP Lagrangian can be written as~\cite{Bauer:2020jbp}
\begin{equation}
\mathcal{L}_a =  \frac{1}{2} (\partial_\mu a )
  (\partial^\mu a ) +\frac{m_a^2}{2} a^2  
  + i  \frac{c_t m_t}{f_a} \bar t
  \gamma_5 \, t  \, a \, ,
    \label{eq:lagr_alp}
\end{equation}
where $a$ is the ALP, and $m_a$ and $f_a$ are the ALP mass and the ALP decay constant, respectively. $c_t$ is the Wilson coefficient parametrising the coupling between the ALP and top quarks. We assume that this coupling is proportional to the top-quark mass, which we accordingly factored out of the Wilson coefficient.\footnote{This coupling structure \TB{can} arise under the assumption that the last term in \cref{eq:lagr_alp} has its origin in the shift-invariant basis 
after electroweak symmetry breaking from operators of the form $\frac{\partial^\mu a}{f_a} \sum_f \bar f \mathbf{c}_f \gamma_\mu f$, with $f = Q_L, u_R, d_R$ (see \GW{e.g.~}\ccite{Anuar:2024myn} for details). Here, we only consider the top-quark operator.}

\begin{figure}
\centering
\includegraphics[width=0.9\textwidth]{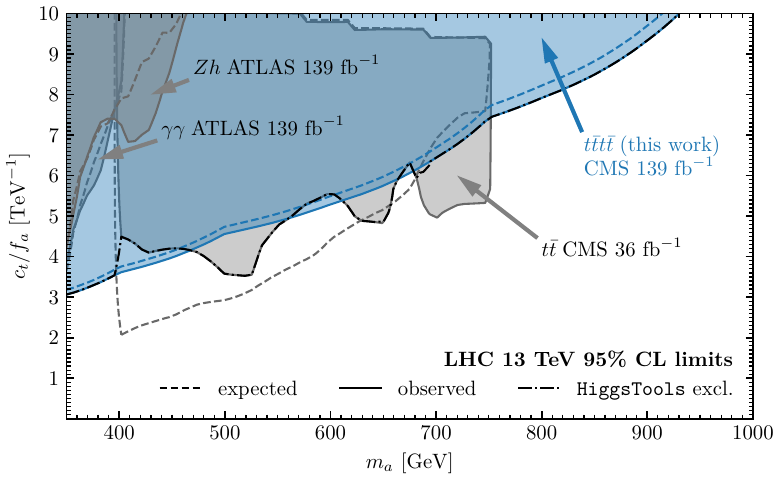}
\caption{95\% CL exclusion regions in the 
\GW{$(m_a , c_t / f_a)$}
plane from our recasting of the CMS searches in final states with four top quarks in blue. The grey regions are excluded from LHC searches for $a \to Zh$~\cite{ATLAS:2022enb}, $a \to \gamma\gamma$~\cite{ATLAS:2021uiz} and $a \to t \bar t$~\cite{CMS:2019pzc}. The solid and dashed lines indicate the exclusions based on the observed and expected cross-section limits, respectively, and the \GW{dot}-dashed line indicates the combined \HiBo exclusion.}
\label{fig:alpplot}
\end{figure}

Assuming that the ALP couples only to top quarks at the scale $m_a$, the ALP can decay at tree level into top-quark pairs, $a \to t \bar t$. At one-loop level, the coupling to top quarks induces the decays $a \to \gamma\gamma$, $a \to Z \gamma$, $a \to Z h$ (we here assume that the detected 125~GeV Higgs boson $h$ behaves as predicted in the SM) and $a \to gg$. In order to 
\GW{investigate} the experimental sensitivity of the LHC to such a top-philic ALP, we show in \reffi{fig:alpplot} the exclusion regions in the plane of the ALP mass $m_a$ vs.~the coupling strength $c_t / f_a$. The exclusions resulting from our recasting of the CMS $\tttt$ searches are shown in blue, whereas the exclusions from other searches \TB{currently implemented in \texttt{HiggsBounds} that are} sensitive to the considered ALP masses and coupling strengths are shown in grey. One can see that the $\tttt$ searches increase the experimental coverage to smaller values of $c_t / f_a$ at and below the di-top threshold $m_a \lesssim 400\gev$ in comparison to the previously most sensitive searches exploiting the $a \to \gamma\gamma$ and $a \to Zh$ decay modes~\cite{ATLAS:2021uiz,ATLAS:2022enb}. At larger ALP masses, our recasting of the $\tttt$ searches increases the experimental sensitivity compared to the searches in the di-top final state, which, however, are based only on the first-year Run~2 data~\cite{CMS:2019pzc}.\footnote{Recently, both ATLAS~\cite{ATLAS:2024vxm} and CMS~\cite{CMS:2025dzq} have published new limits taking into account the full Run~2 data. These results will be included in a future release of the \HiBo dataset.
} It should also be noted that the largest values of $m_a$ considered here are 
close to the scale $\Lambda = 4 \pi f_a$
(assuming $c_t \lesssim 1$).
\TB{This scale represents} 
the cutoff of the ALP effective field theory where operators with dimension larger than five become relevant.

\paragraph{A heavy \texorpdfstring{\cp}{CP}-mixed Higgs boson}

\begin{figure}[!htpb]
\centering
\includegraphics[width=0.6\textwidth]{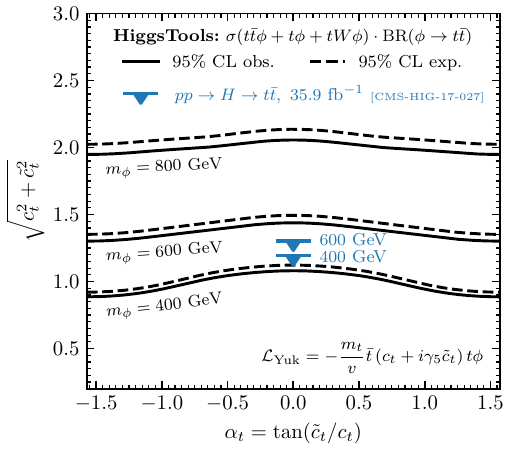}
\caption{
Upper expected (dashed) and observed (solid)
limits on the coupling strength
$\sqrt{c_t^2 + \tilde c_t^2}$ as a function of
the angle $\alpha_t$ for a \cp-mixed Higgs boson
$\phi$ for three different masses
$m_\phi = 400,600,800\gev$ resulting
from our recasting of the CMS searches in final
states with four top quarks~\cite{CMS:2019rvj}.
Also shown with blue markers are the corresponding
observed upper limits from CMS searches in the
$t \bar t$ final state~\cite{CMS:2019pzc},
which are applicable for a \cp-even Higgs boson, i.e.~at $\alpha_t = 0$.}
\label{fig:cpvioplot}
\end{figure}

As a third example, we consider a heavy \cp-mixed Higgs boson $\phi$ with couplings as it could for example appear in the \cp-violating complex 2HDM~\cite{TDLee}.
In \cref{fig:cpvioplot}, we show limits on \GW{the coupling strength of} $\phi$ as a function of $\alpha_t = \tan(\tilde c_t/c_t)$ for three different mass values. While the \GW{indicated di-top limit 
from CMS~\cite{CMS:2019pzc}}
is only applicable for a purely \cp-even (
\KR{alternatively, also the pure \cp-odd limit is provided~\cite{CMS:2019pzc}}) 
scalar, the newly implemented 
limit \GW{from our recasting} provides bounds also in \cp-admixed scenarios, for example limiting the combined coupling strength $\sqrt{c_t^2 + \tilde c_t^2}$ to be lower than $\sim 1.4$ for $m_\phi = 600 \gev$ 
\TB{across all possible values of $\alpha_t$.}


\subsection{Coupling-dependent non-resonant di-Higgs limits}

A further 
feature of the new version of \HiBo includes the possibility to define coupling-dependent pair-production limits.
This is a useful feature in particular for the case of \KR{the $h$-pair} 
production \GW{process}, for which the
\TB{experimental sensitivity, and thus the}
exclusion limits, depend on the trilinear Higgs coupling $\kappa_{\lambda}$. 
%
The \TB{new \texttt{HiggsBounds}} implementation
\TB{of the existing cross-section limits from non-resonant Higgs-boson
pair production}
is based on an additional acceptance factor that captures 
\GW{the} dependence on the trilinear Higgs coupling
when defining the exclusion limit. To determine the acceptance factor, we approximate the observed exclusion limit as a function of $\kappa_\lambda$ using the rational expression
\begin{equation}
    \sigma_{\mathrm{obs}} = \frac{A\kappa_\lambda^2+B\kappa_\lambda+C}{D\kappa_\lambda^2+E\kappa_\lambda+F}\sigma_{\mathrm{SM}},
    \label{eq:noresacc}
\end{equation}
\KR{where $\sigma_{\rm SM}$ is the \GW{predicted value in the SM} 
assumed in the corresponding experimental analysis.}
The coefficients $A$, $B$, $C$, $D$, $E$, and $F$ are obtained via a fit to the observed experimental exclusion limit using FindFit in
\texttt{Mathematica}~\cite{Mathematica}. 
These coefficients along with the value of the limit at $\sigma_{\mathrm{obs}}(\kappa_{\lambda} = 1)$ are implemented in the 
limit definition files for each implemented \TB{search for} non-resonant 
\TB{Higgs-boson pair production}. Because combinations of multiple decay channels yield 
significantly stronger constraints than individual channels, we include in the current implementation the two 
ATLAS~\cite{ATLAS:2024ish} 
and CMS~\cite{CMS:2025ngq}
\TB{combinations}\KR{, as well as the recent combination of both experiments, which is so far only a preliminary result~\cite{ATLAS-CONF-2025-012}}.\footnote{\TB{To} 
ensure the \TB{correct} 
application of the combined limits, we demand the 125~GeV scalar to have SM-like branching ratios.} \KR{Additionally, we include the first Run 3 non-resonant analysis by ATLAS in the $b\bar{b}\gamma\gamma$ final state~\cite{ATLAS:2025hhd}, which has comparable limits to the Run 2 combinations.}
We provide an example of the implementation \KR{of the latter limit} as a JSON file in \cref{app:kala_dep_limits}.

\begin{figure}[!htpb]

\centering
\includegraphics[trim={.1cm .1cm .1cm .1cm},clip,width=0.44\textwidth]{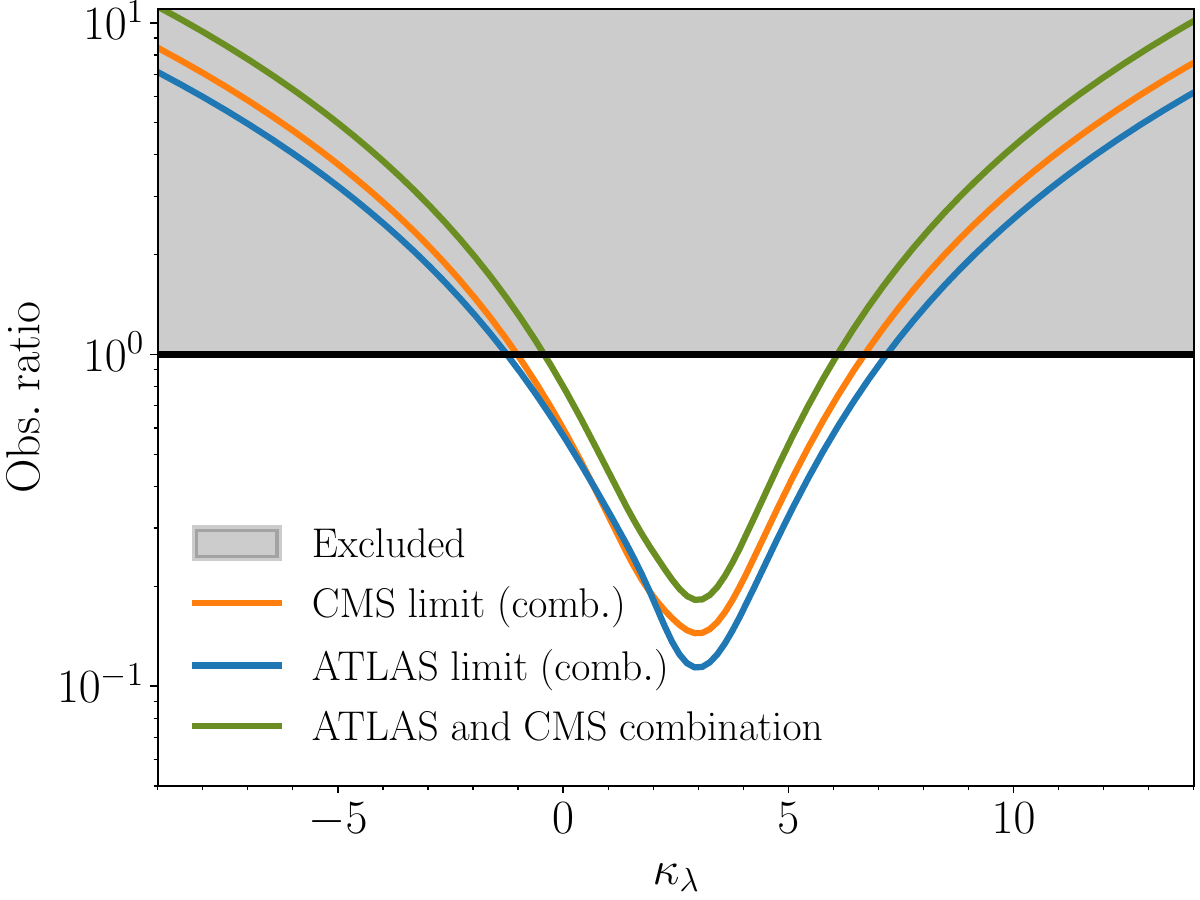}~
\includegraphics[trim={.1cm .1cm .1cm .1cm},clip,width=0.44\textwidth]{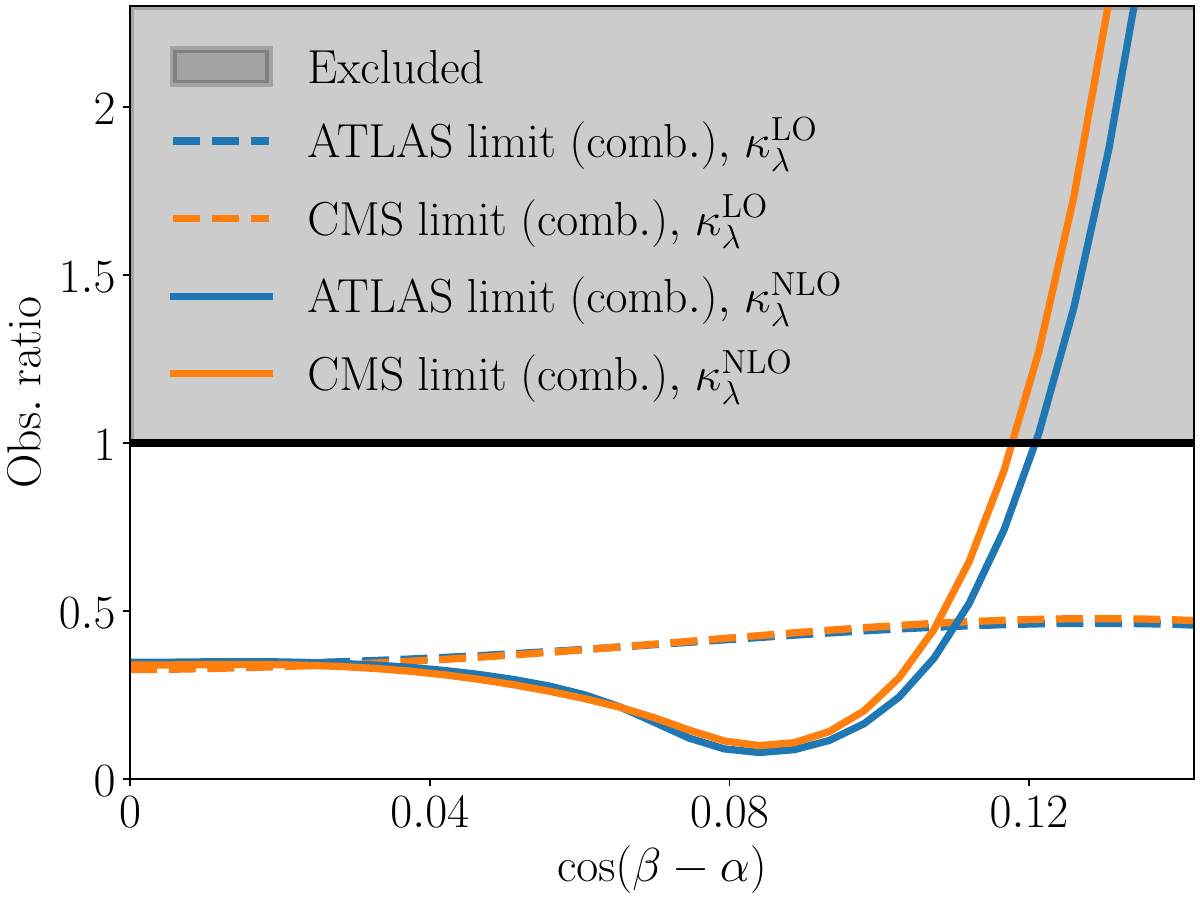}
\caption{Coupling-dependent non-resonant di-Higgs limits. Left: observed ratio for an effective value of $\kappa_{\lambda}$. 
Right: observed ratio for different values of $\cos(\beta-\alpha)$ in \GW{a} Type~I~2HDM depending on the \GW{predicted} value of $\kappa_{\lambda}$ \GW{in this model, for which the cases are shown where the}
tree-level (dashed) \GW{and the} one-loop (solid)
\GW{predictions are used}. For the other parameters, see text.}
\label{fig:kala_dep_limits}
\end{figure}

In the left panel \TB{of \cref{fig:kala_dep_limits}}, 
we show the ratio of the predicted 
\GW{cross-section 
as a function of
\KR{$\kappa_\lambda$} 
divided by} the observed cross-section limit for 
\GW{this value of $\kappa_\lambda$}
according to the results provided by CMS~\cite{CMS:2025ngq} 
(orange line) and ATLAS~\cite{ATLAS:2024ish} (blue line) combinations\KR{, as well as the combination of both experiments from full Run 2~\cite{ATLAS-CONF-2025-012}} \GW{(green line)}. 
The 
\GW{grey} shaded area shows the \GW{excluded} region 
\GW{where this ratio is} larger than one.
\GW{Currently the ATLAS and CMS combination based on the full Run 2 data~\cite{ATLAS-CONF-2025-012} yields the strongest bounds on} 
$\kappa_{\lambda}$.


The right plot \TB{of \cref{fig:kala_dep_limits}}
illustrates the application of the coupling-dependent limits 
\GW{for a particular}
BSM scenario.
For this example, we chose the 2HDM Type I, though this procedure is general and can be used for any model that predicts a deviation of
$\kappa_\lambda$ from the SM \GW{case}. 
We use a benchmark scenario with a negligible resonant contribution, 
with $\tan\beta =10$, $m_H=m_A=m_{H^{\pm}} = 1000 \gev$ and $m_{12}^2=m_H^2c_{\alpha}/\tan\beta$. 
\TB{The limits from non-resonant Higgs-boson pair production
can be readily}
applied by adding $\kappa_\lambda$ to the \texttt{effectiveCouplingInput} feature of \HiPr.
\GW{In this case we show the impact of the limits from the 
ATLAS combination~\cite{ATLAS:2024ish} (blue lines)
and from the CMS combination~\cite{CMS:2025ngq} 
(orange lines). 
The plot shows the results for the observed ratio as function of 
$\cos(\beta-\alpha)$ for the two cases where the}
tree-level (dashed line) \TB{and} 
\GW{the} 
full one-loop corrections (solid line) 
\GW{for the prediction of $\kappa_\lambda$ are used}, the latter computed using 
\texttt{anyH3}~\cite{Bahl:2023eau}.\footnote{\TB{We assume here
that the dominant EW NLO corrections to the cross-section
result from the radiative corrections to the
trilinear Higgs-boson self coupling, which is
a good approximation for large positive values of
$\kappa_\lambda \sim 6$ that can currently
be probed at the LHC~\cite{Bahl:2022jnx}.}}
\GW{Incorporating the loop corrections into the prediction for 
$\kappa_\lambda$ is seen to have a big effect~\cite{Heinemeyer:2024hxa}.
While for the tree-level prediction (dashed lines) the observed ratio stays far below one for all displayed values of $\cos(\beta-\alpha)$, 
incorporating the loop contributions into the prediction for $\kappa_\lambda$ drastically changes the observed ratio and gives rise to an excluded region for $\cos(\beta-\alpha)$. According to the standard procedure in \HiBo the experimental result that is applied for setting the limit is the one that}
corresponds to the most sensitive channel, i.e.\ the one that maximises 
the ratio between the model prediction and the expected 95\% C.L.~exclusion bound on the cross-section. 
In this \GW{example}, that channel is the CMS combination.
\GW{It yields for the case where the}
loop-level corrections to $\kappa_\lambda$ \GW{are incorporated an} exclusion for 
\GW{$\cos(\beta-\alpha) > 0.11\KR{8}$ 
(the result from the ATLAS combination would have given rise to the limit
$\cos(\beta-\alpha) > 0.12$).} 
\KR{These results are in agreement with the ones presented in Ref.~\cite{Heinemeyer:2024hxa}.}

The current implementation benefits from the flexibility of the \HiTo\ framework: once the effective couplings of a BSM point are specified, 
all relevant limits are automatically considered. 
As a result, non-resonant di-Higgs searches are now integrated into global analyses of BSM scenarios.


\subsection{Improved handling of pre-Higgs discovery searches}

\HiBo is not able to combine experimental searches, because the necessary information for \GW{performing} a combination is typically not public. Therefore, \GW{as mentioned above}, only the most sensitive search is selected based on the expected limit. For the selected search, the observed limit is applied. In this way, a strict 95\% confidence level interpretation is maintained.

\HiBo includes pre-Higgs discovery searches for which --- due to the 
\GW{manifestation of the} later discovered Higgs boson 
\GW{in the data} --- the expected limit is much stronger than the observed one. In previous versions, since \HiBo always selects the limit to evaluate based on the expected limit, for SM-like Higgs bosons with a mass close to 125~GeV, nearly always one of the pre-Higgs discovery searches 
\GW{was}
selected. Consequently, other searches --- e.g., for non-resonant di-Higgs production or for invisible decays of the 125 GeV Higgs boson --- 
\GW{were} only rarely selected even though they would 
\GW{actually}
exclude the specified parameter point. 

To avoid this kind of behaviour, one could think of removing the pre-Higgs discovery limits. They, however, cover a broad mass range around $125 \gev$ and are not all superseded by more recent searches \TB{with more stringent limits outside
of the \GW{region where a} 
resonance \GW{peak} at about $125\gev$ \GW{occurs}.
Therefore, i}nstead of removing 
\TB{the pre-Higgs discovery searches}
completely, \HiBo now offers a switch to suppress their selection if the scalar mass falls within a certain mass window, which we choose to be 125 GeV plus/minus the experimental resolution.
\TB{Outside of the mass window around the 125~GeV
\GW{resonance}, the cross-section limits of the
pre-Higgs discovery searches are therefore still applied
to light BSM Higgs bosons with masses in the
vicinity of 125~GeV.}

In practice, this functionality can be
\TB{selected or turned off}
by supplying an additional optional argument to the \HiBo object:
\begin{minted}[bgcolor=bg]{python}
import Higgs.predictions as HP
import Higgs.bounds as HB

pred = HP.Predictions() # create the model predictions object
bounds = HB.Bounds('/Path/To/HBDataSet') # load HB dataset

# add particles and provide predictions
# ...

# HB result with suppression of pre-Higgs discovery limits
res1 = bounds(pred, suppressLimits=True)
# HB result without suppression of pre-Higgs discovery limits
res2 = bounds(pred, suppressLimits=False)
\end{minted}
If \TB{the optional 
argument \texttt{suppressLimits} is not} supplied, the selection of the pre-Higgs discovery limit is suppressed\TB{,
i.e.~the default value of \texttt{suppressLimits} is
\texttt{True}.}


\section{HiggsSignals}
\label{sec:hs}

In this Section, we present several improvements of \HiSi\ and discuss the proper handling of the evaluated $\chi^2$.


\subsection{Handling scalars with mass uncertainties}
\label{sec:massunc}

Scalars with mass uncertainties are a special case, which e.g.\ appears in supersymmetric theories~\cite{Slavich:2020zjv}.
While \HiTo allows straightforwardly to input mass uncertainties for each scalar, the computation of the model predictions requires special care by the user.

This is, in particular, the case if the predicted central mass value $m$ for the $h$ candidate scalar deviates sizeably from $125\gev$ within the mass uncertainty $\Delta m$ (e.g., if $m + \Delta m \simeq 125\gev$ with $\Delta m =2\gev$). While the mass measurements are not able to exclude such a point \GW{in view of the theoretical uncertainty of the mass prediction}, computing the model predictions for the cross-section and branching ratios at the mass $m$ (and not $125\gev$) can lead to 
\GW{large}
deviations from the data.
\GW{This can be the case}
even if all couplings are
\TB{predicted to be} SM-like \TB{in the model under investigation} 
\GW{as a consequence of the sensitive}
mass dependence of the cross-sections and branching ratios (in particular of the $H\to WW^*$ and $H\to ZZ^*$ partial widths). 

To avoid this issue, we strongly recommend calculating not only the BSM prediction at the mass $m$ but also the corresponding SM prediction at the same mass including the same level of higher-order corrections. Then, the ratio of both should be formed and multiplied by the 
\GW{state-of-the-art}
SM prediction for a mass of $125 \gev$. \HiTo handles this rescaling automatically if the effective coupling input is used (see also \cref{sec:rescaling} \GW{below}).
It should be noted, however, that this procedure is not able to handle BSM decay modes. To pass information about BSM decay modes, the respective partial widths can be given after
the effective coupling input
\TB{has been used to obtain the partial widths
for decays into SM particles}. To facilitate the handling of these issues, we \GW{have} extended the SLHA interface with optional arguments which allow specifying which branching ratios should be calculated in terms of the 
\GW{provided}
effective couplings and which should be directly taken from the SLHA file (see the online documentation for further details).




\subsection{Scaling to reference mass of measurements}
\label{sec:rescaling}

We 
\GW{now turn to}
the scaling of predictions to the reference mass of signal strength measurements. Each measurement of a signal strength, $\mu^\text{obs} \equiv \sigma^\text{obs}/\sigma_\text{ref}^\text{obs}(m_\text{ref})$, depends on the reference mass $m_\text{ref}$ for which the denominator reference cross-section --- typically the SM --- is evaluated 
\GW{(for simplicity, we use here the phrase cross-section and the symbol $\sigma$, but it is understood that the signal strength is obtained from the production cross-section times the decay branching ratio, normalised to the appropriate reference value)}.
If we compare such a signal strength measurement to predictions within \HiSi, we have to take into account that \GW{in the predictions} the $h$ candidate might have a mass $m$
\TB{that is slightly different from the
reference mass,} $m \neq m_\text{ref}$.\footnote{In fact, different measurements use (slightly) different reference masses. \GW{Thus}, always choosing $m = m_\text{ref}$ is impossible.}

In practice, the predicted and observed rates are compared. This means that we have to compute the model prediction for the rate, which is evaluated from the predicted signal strength $\mu$ and the reference rate of the observable as
\begin{align}
    \mu \cdot \sigma^\mathrm{obs}_\mathrm{ref}(m_\mathrm{ref}) =
    \frac{\sigma^\mathrm{incl}(m)}{\sigma^\mathrm{incl}_\mathrm{ref}(m_\mathrm{norm})}
    \sigma^\mathrm{obs}_\mathrm{ref}(m_\mathrm{ref})\,,
    \label{eq:musigmaobs}
\end{align}
where $m$ is the particle mass 
\GW{provided as input by the user}, and
\TB{$\sigma^{\rm incl}(m)$}
is used to denote the inclusive cross-section for a certain channel (e.g., Higgs production via gluon
fusion)\TB{, which is related to the observed cross-section
via the experimental acceptance
\begin{equation}
  \mathcal{A}(m) = \frac{\sigma^{\rm obs}(m)}
    {\sigma^{\rm incl}(m)} \, .
\end{equation}
At this point, the choice arises whether to}
choose $m_\mathrm{norm} = m_\mathrm{ref}$ or $m_\mathrm{norm} = m$. Somewhat counter-intuitively, choosing $m_\mathrm{norm} = m_\mathrm{ref}$ --- i.e., normalizing the signal strength to the reference mass --- means 
\GW{that} the rate 
\GW{is {\em not\/} rescaled}
to the reference mass. 
\GW{This choise} turns
\TB{the right-hand side of \cref{eq:musigmaobs}} into
\begin{align}
    \sigma^\mathrm{incl}(m)
    \frac{\sigma^\mathrm{obs}_\mathrm{ref}(m_\mathrm{ref})}{\sigma^\mathrm{incl}_\mathrm{ref}(m_\mathrm{ref})}
    = \sigma^\mathrm{incl}(m)\mathcal{A}_\mathrm{ref}(m_\text{ref})\,,
    \label{eq:norescaling}
\end{align}
which assumes the reference acceptance $\mathcal{A}_\text{ref}$ to be constant over the mass range $[m_\mathrm{ref},m]$, and directly uses the model-predicted
\TB{cross-section $\sigma^{\rm incl}(m)$}
without any rescaling. Choosing $m_\mathrm{norm} = m$ --- i.e., using the signal strength normalised at the particle mass
\TB{$m$ that is provided as input by the user} --- instead leads to 
\begin{align}
\sigma^{\rm incl}(m) \frac{\sigma^{\rm obs}_{\rm ref}(m_{\rm ref})}{\sigma^{\rm incl}_{\rm ref}(m)} &=
  \sigma^{\rm incl}(m)
  \frac{\sigma^{\rm incl}_{\rm ref}(m_{\rm ref})}{\sigma^{\rm incl}_{\rm ref}(m)}
  \frac{\sigma^{\rm obs}_{\rm ref}(m_{\rm ref})}{\sigma^{\rm incl}_{\rm ref}(m_{\rm ref})} =
  \sigma^{\rm incl}(m) \cdot f \cdot \mathcal{A}_\mathrm{ref}(m_\mathrm{ref}) \, ,
  \label{eq:dorescaling}
\end{align}
which \TB{effectively} rescales the model-predicted inclusive rate to the reference mass by the factor
\TB{\begin{equation}
f = \frac{\sigma^{\rm incl}_{\rm ref}(m_{\rm ref})}
  {\sigma^{\rm incl}_{\rm ref}(m)}\;,
\end{equation}
where it is assumed}
that the mass dependence of the model rate matches the mass dependence of the reference rate over the mass range $[m_\mathrm{ref},m]$.
\TB{In this way},
\HiSi\ \TB{effectively} approximates what the model-predicted rate would be if the particle mass were exactly equal to the reference mass. For inclusive rates that are the sum of multiple channels, this scaling is performed separately for each channel.

\HiSi offers multiple options to control this rescaling:
\begin{itemize}
    \item[1.] always directly use the model-predicted rates
    \TB{as shown in \cref{eq:norescaling}}
    (no rescaling);
    \item[2.] always rescale the model-predicted rates to the reference mass
    \TB{as shown in \cref{eq:dorescaling}};
    \item[3.] only rescale the model-predicted rate to the reference mass if the reference mass lies within the theoretical mass uncertainty
    \TB{$\Delta m$} of the particle
    \TB{(see the discussion in \cref{sec:massunc})}.
\end{itemize}
The \TB{new} default option is to always rescale
the model-predicted rates to the reference mass
\TB{(option 2), irrespectively of the user input
for $m$ and $\Delta m$}.
In earlier versions of \HiSi, the rescaling within the mass uncertainty of the particle has been used
\TB{(option 3). However, this}
can lead to
\TB{a spurious behaviour in which the fit to the data becomes
worse 
with increasing mass uncertainty of the particle.}



\subsection{Recommendation for using the \texorpdfstring{\HiSi}{HiggsSignals} \texorpdfstring{\boldmath{$\chi^2$}}{chi2} output}
\label{sec:chisqrecom}

\HiSi provides a total chi-squared value $\chi^2$
that quantifies how well a given model point reproduces the
included set of Higgs boson measurements.
This value accounts for both statistical and systematic
uncertainties, as well as known correlations between the
measurements, and serves as a global test statistic for
the compatibility between theory and data.
In most of the use cases the goal for users is to
determine whether a specific parameter point of
a model can be excluded at a given confidence level,
usually at 95\%.
One approach is to treat the total $\chi^2$ as a test
statistic that follows a chi-squared distribution with
$\nu = n- n_p$ degrees of freedom, where $n$ is the number of
measurements and $n_p$ is the number of free parameters
in the model. However, this assumption is problematic
and leads to incorrect statistical interpretations
in the context of \HiSi.
First, the individual measurements are often
significantly correlated, reducing the effective number
of independent data points.
Second, not all free parameters of a model
impact the Higgs boson observables equally. Some may only
weakly affect the fit, or only under specific conditions,
while others may be entirely unconstrained by the data.
In practice,
these issues 
\GW{give rise to the fact that using such a prescription based on the total $\chi^2$ will often have the result that}
parameter points 
\GW{are classified as being not excluded at the given confidence level}
even
\GW{if} they provide a poor description of the data,
notably because $n \approx 150$ is large.
This makes exclusion criteria based on $\chi^2$
thresholds derived from $n - n_p$ problematic for
identifying allowed parameter points or parameter regions.

A statistically more meaningful approach is to use
\GW{a suitable}
difference in $\chi^2$ values, e.g.
\begin{equation}
    \Delta \chi^2 = \chi^2 - \chi^2_{\rm best-fit} \, ,
\end{equation}
and compare it to a 
\GW{chosen} threshold value.\footnote{If
it is impractical to obtain the value
$\chi^2_{\rm best-fit}$ in a certain model, for instance because the model
has many free parameters,
an alternative choice is to use the value predicted
by the SM, $\chi^2_{\rm SM}$, instead, as long as the
experimental data 
\GW{is sufficiently} well compatible with the SM such
that $\chi^2_{\rm SM}$ can be considered \GW{as} a good approximation
of $\chi^2_{\rm best-fit}$.}
This approach focuses on how much worse 
\GW{the $\chi^2$ of}
a parameter
point of a model is
compared to the best-fit point in the considered
parameter space and
avoids the need to estimate an effective total number
of degrees of freedom $\nu$.
For example, in a two-dimensional parameter scan,
a common and \GW{well} justified criterion is to exclude points where
$\Delta \chi^2 > 5.99$, corresponding to a 95\% confidence
level for two degrees of freedom.
This is especially suitable \GW{if} only two parameters are
varied in the scan, \GW{while} the remaining ones are either fixed
or properly profiled.
Even in broader scans involving more free parameters,
where multiple parameters are varied without profiling,
applying the $\Delta \chi^2$ method with a low number
of degrees of freedom (e.g.~$\nu = 1,2$)
can still be useful. 
While the formal confidence level associated with a fixed $\Delta \chi^2$ threshold assumes that only a small number of parameters are varied (or the others are properly profiled), the method remains a useful diagnostic in broader scans. In such cases, it provides a consistent relative criterion for identifying worse-fitting parameter points, even though it may not 
\GW{have a direct interpretation as a}
statistical confidence level.
In this way, the $\Delta \chi^2$-based approach
quantifies the relative quality of the fit,
rather than relying on absolute $p$-values, 
which can be misleading \GW{if} $n \gg n_p$, 
thus remaining the preferable option.
Users are encouraged to adopt the $\Delta \chi^2$-based
exclusion method, using thresholds corresponding to one
or two degrees of freedom depending on the dimensionality
and interpretation of their scans, especially when seeking
to 
\GW{exclude}
parameter points that are clearly disfavoured by
the data on a point-by-point basis.

It 
\GW{should be mentioned}
that, if \GW{technically} possible, profiling
(of additional nuisance parameters) in a
likelihood-based inference 
approach
or empirical calibration in order to determine
the actual $\chi^2$ distribution (e.g.~via Monte-Carlo
sampling) are superior methods that can further
improve the reliability of the statistical interpretation
of the fit. 
With the total $\chi^2$ value provided by
\HiSi, users are free to apply more sophisticated statistical
methods such as the ones just mentioned, according to their 
specific goals. As a model-agnostic tool, \HiSi
deliberately leaves such interpretation steps to the user.

For a more in-depth discussion, see \ccite{Bahl:2022igd}.


\section{Conclusions}
\label{sec:conclusions}

We have presented a new version \texttt{1.3} of the public computer program \texttt{HiggsTools}, which overarches
the three codes \texttt{\HiPr-2.0}, \texttt{\HiBo-6.1} and \texttt{\HiSi-3.1}.
\HiPr is a joint interface for the predictions of the various Higgs production
and decay rates, facilitates the definition
of the physical model and \GW{provides} the input for \HiBo and \HiSi.
The code \HiBo tests 
\GW{arbitrary}
BSM models against exclusion
limits from LEP and LHC Higgs searches. \HiSi confronts
the predictions of arbitrary BSM models with the measured mass and rates of the Higgs
boson that has been detected at about 125 GeV at the LHC. 

The description of \HiTo and its sub-programs \HiPr, \HiBo and \HiSi provided in the present 
paper has focused on the improvements of the functionality and applicability of the codes 
with respect to the previous versions as presented in \ccite{Bahl:2022igd}.
For \HiPr, the user has to specify the scalar sector of the model under
consideration: the properties of each scalar, including the mass, total width,
charge, \cp-character and the rates for all relevant production and decay channels. These
properties can be set by the user directly, or alternatively via the effective couplings of the scalars.
W.r.t.\ the previous version, three major improvements have been included: (i) \HiPr now provides in addition 
to the cross-sections at $13 \tev$ also the cross-sections at $13.6 \tev$ and $14 \tev$ for scalar masses 
up to $1000$, $3000$, or $5000 \gev$, depending on the production channel.  
(ii) For the gluon fusion cross-section, two options were implemented:
\texttt{SMHiggsEW} including N3LO QCD corrections in the infinite top \TB{quark} mass limit as well as NLO
electroweak corrections, suitable for lower scalar masses; 
\texttt{SMHiggs} including NNLO QCD corrections for a finite top \TB{quark} mass, suitable for higher scalar masses.
\HiPr now provides a smooth interpolation between the two, referred to as \texttt{SMHiggsInterp}.
(iii) \HiPr now provides cross-section predictions for $hh$ production (where $h$ denotes the
Higgs \GW{boson} discovered at the LHC).
In the non-resonant case, the prediction requires $\kappa_\lambda \equiv \lambda_{hhh}/\lambda_{hhh}^{\mathrm{SM}, \GW{(0)}}$
as input. For 
resonant di-Higgs production, i.e.\ in the case of a contribution of a heavy \cp-even Higgs boson 
via
$gg \to H \to hh$, additionally the mass and the width of the heavy Higgs \TB{boson} \GW{have} to be provided,
together with its top\TB{-quark} Yukawa coupling and the relevant trilinear Higgs coupling, $\lambda_{hhH}$. 

Concerning \HiBo, also three major improvements have been incorporated into the new version.
(i) 
While di-top searches had already been implemented in \HiBo, 
now also \TB{BSM scalar} searches \TB{in} 
final states with three or four top quarks are available, 
based on a recasting of a corresponding CMS limit. These multi-top final states 
occur in the 
$t\bar t \phi$, $t\phi$ and $tW\phi$ production modes with a subsequent decay $\phi \to t\bar t$. 
We demonstrated the relevance of these searches for the 2HDM, a top-philic ALP, and a heavy
\cp-mixed Higgs boson. Particularly for the latter, the 
experimental limits \GW{published so far by ATLAS and CMS} are either
for a \cp-even or a \cp-odd Higgs boson, while \HiBo also covers the \cp-mixed case.
(ii) Searches for non-resonant di-Higgs production have been included. Since the prediction of the di-Higgs 
cross-section is based on the input of $\kappa_\lambda$, \HiBo facilitates to set limits on this parameter  
in BSM models. 
(iii) \HiBo includes pre-Higgs discovery searches for which, due to the later discovered Higgs
boson, the expected limit is much stronger than the observed one. 
For SM-like Higgs bosons with a mass close to 125 GeV, in previous \HiBo versions
nearly always one of the pre-Higgs discovery searches was selected 
\GW{as the most sensitive channel}. As a new default, these
searches are 
not applied for Higgs bosons with a mass 
\GW{that is compatible with
$125 \gev$ 
within}
the experimental resolution. This 
\GW{setting}
can, however, be overwritten by the user.

For \HiSi, \GW{in particular two} 
improvements have been included into the new version:
(i) The handling of mass uncertainties particularly in cases 
\GW{where relatively} large deviations in the
Higgs-boson mass prediction 
\GW{from the experimental value can occur}
(as in supersymmetric theories) has been improved. 
(ii) Each signal strength measurement is given at a certain reference mass, often not coinciding with 
the mass of the SM-like Higgs boson under consideration. 
In the new version of \HiSi the model prediction, e.g.\ for a signal strength, is rescaled from 
the input mass provided by the user to the reference mass used in the experimental measurement.
In other words, \HiSi approximates what the model-predicted
rate would be if the particle mass were exactly equal to the reference mass, leading to a more
reliable application of the experimental measurements.
We have furthermore provided a short recommendation for using the \HiSi $\chi^2$ facilitating the identification of parameter points that are clearly disfavoured by the data on a point-by-point
basis.

\GW{With these new features and improvements and the \texttt{HiggsTools} framework should be well prepared for exploiting the experimental results from Run~3 of the LHC and beyond.}

\bigskip
The code \texttt{HiggsTools-1.3}, containing \HiPr, \HiBo and \HiSi, is available via
\begin{center}
\texttt{https://gitlab.com/higgsbounds/higgstools}\,.
\end{center}


\section*{Acknowledgements}
\sloppy{
We thank Cheng Li and Steven Paasch for their collaboration in the early stages of this work. Moreover, we thank Benjamin Fuks for helpful discussions. 
H.B.~acknowledges support through the KISS consortium (05D2022) funded by the German Federal Ministry of Education and Research BMBF in the ErUM-Data action plan, by the Deutsche Forschungsgemeinschaft (DFG, German Research Foundation) under grant
396021762 – TRR 257: Particle Physics Phenomenology after the Higgs Discovery, and through Germany’s Excellence Strategy EXC 2181/1 – 390900948 (the Heidelberg STRUCTURES Excellence Cluster) and by the state of Baden-Württemberg through bwHPC and the German Research Foundation (DFG) through grant INST 35/1597-1 FUGG.
G.W.\ acknowledges support by the Deutsche Forschungsgemeinschaft (DFG, German Research Foundation) under Germany's Excellence Strategy -- EXC 2121 ``Quantum Universe'' – 390833306. This work has been partially funded by the Deutsche Forschungsgemeinschaft (DFG, German Research Foundation) - 491245950.
The work of S.H.\ has received financial support from the
grant PID2019-110058GB-C21 funded by
MCIN/AEI/10.13039/501100011033 and by ``ERDF A way of making Europe''. 
S.H.\ also acknowledges support from Grant PID2022-142545NB-C21 funded by
MCIN/AEI/10.13039/501100011033/ FEDER, UE.
T.B., S.H.\ and K.R.\ acknowledge support
of the Spanish Agencia
Estatal de Investigaci\'on through
the grant IFT Centro de Excelencia Severo Ochoa CEX2020-001007-S
funded by MCIN/AEI/10.13039/501100011033. 
The project that gave rise to these
results received the support of a
fellowship from the ``la Caixa''
Foundation (ID 100010434). The
fellowship code is LCF/BQ/PI24/12040018.
}


\appendix


\section{Additional information on \GW{the} implementation of \GW{the} CMS four-top limit}
\label{app:HB}

\input{app_HB.tex}


\bibliographystyle{JHEP}
\bibliography{bibliography}

\end{document}

%% file: abbr.tex
\newcommand{\GW}[1]{{\color{black}#1}}

\newcommand{\TB}[1]{{\color{black} #1}}

\newcommand{\KR}[1]{{\color{black}#1}}

\newcommand{\SH}[1]{{\color{black}#1}}

\newcommand{\SHd}[1]{\SH{\st{#1}}}

\newcommand{\cp}{\ensuremath{{\cal CP}}\xspace}

\newcommand{\tttt}{\mbox{$t \bar t t \bar t$}\xspace}

\newcommand\pdfmath[1]{\texorpdfstring{$#1$}{#1}}

\newcommand{\cpp}{\texttt{C++}\xspace}
\newcommand{\py}{\texttt{Python}\xspace}
\newcommand{\mat}{\texttt{Mathematica}\xspace}

\newcommand{\gev}{\;\text{GeV}\xspace}
\newcommand{\tev}{\;\text{TeV}\xspace}

\newcommand{\HiTo}{{\texttt{HiggsTools}}\xspace}
\newcommand{\HiPr}{{\texttt{HiggsPredictions}}\xspace}
\newcommand{\HiBo}{{\texttt{HiggsBounds}}\xspace}
\newcommand{\HiSi}{{\texttt{HiggsSignals}}\xspace}

\newcommand\Code[1]{\ensuremath{\texttt{#1}}}

\def\reffi#1{\mbox{Fig.~\ref{#1}}}

\newcommand\refse[1]{Sect.~\ref{#1}}

\newcommand\ccite[1]{Ref.~\cite{#1}}

%% file: app_HB.tex
\subsection{Cross-section coefficient fit functions}
\label{app:tttt_fit_functions}

In the following we show all fit functions $c_i(m_{\phi})$, where $i \in \{1,7\}$, used to calculate the \GW{cross-sections} depending on the mass $m_{\phi}$ and the couplings $c_t,\tilde c_t$ and $c_V$ as described in \SH{\refse{sec:recast}}.


\subsubsection{\texorpdfstring{$t\bar t\phi$}{ttphi} production}

The fit functions for the coefficients in $\sigma_{\epsilon,t\bar t\phi}$ are given in the form $c_{i,\epsilon, ttH}$ by
\begin{align}
    \begin{split}
        c_{5,\epsilon,t\bar t\phi}(m_{\phi}) &= 2523.55 \cdot \frac{1}{m_{\phi}^2} - 0.0000079 \cdot m_{\phi} + 0.0076 ,\\
        c_{6,\epsilon,t\bar t\phi}(m_{\phi}) &= 0 ,\\
        c_{7,\epsilon,t\bar t\phi}(m_{\phi}) &= 6087.51 \cdot \frac{1}{m_{\phi}^2} + 0.00000027 \cdot m_{\phi} - 0.0039.
    \end{split}
\end{align}
This gives the function for $\sigma_{\epsilon,t\bar t\phi}$ as
\begin{equation}
    \sigma_{\epsilon,t\bar t\phi}(c_t,c_{\tilde{t}}, c_V, m_{\phi}) = c_{5,\epsilon,t\bar t\phi}(m_{\phi}) \cdot c_t^4 +  c_{7,\epsilon,t\bar t\phi}(m_{\phi}) \cdot \tilde{c}_{t}^4 .
\end{equation}
The fit functions for the total cross section $\sigma_{\text{tot},t\bar t\phi}$ are given by 
\begin{align}
    \begin{split}
        c_{5,\text{tot},t\bar t\phi}(m_{\phi}) &= 2703921.1 \cdot \frac{1}{m_{\phi}^2} + 0.0019 \cdot m_{\phi} - 3.71 ,\\ 
        c_{6,\text{tot},t\bar t\phi}(m_{\phi}) &= 0 ,\\
        c_{7,\text{tot},t\bar t\phi}(m_{\phi}) &= 4300877.59 \cdot \frac{1}{m_{\phi}^2} + 0.0048 \cdot m_{\phi} - 8.14.
    \end{split}
\end{align}
This gives the function for $\sigma_{\text{tot},t\bar t\phi}$,
\begin{equation}
    \sigma_{\text{tot},t\bar t\phi}(c_t,c_{\tilde{t}}, c_V, m_{\phi}) =  c_{5,\text{tot},t\bar t\phi}(m_{\phi}) \cdot c_t^4 +  c_{7,\text{tot},t\bar t\phi}(m_{\phi}) \cdot \tilde{c}_{t}^4 \;.
\end{equation}



\subsubsection{\texorpdfstring{$tW\phi$}{tWphi} production}

The fit functions for the coefficients in $\sigma_{\epsilon,tWH}$ are given by
\begin{align}
    \begin{split}
        c_{1,\epsilon,tW\phi}(m_{\phi}) &= 850 \cdot \frac{1}{m_{\phi}^2} - 0.00000042 \cdot m_{\phi} - 0.00018 \;, \\
        c_{2,\epsilon,tW\phi}(m_{\phi}) &= 0 \;, \\
        c_{3,\epsilon,tW\phi}(m_{\phi}) &= - 2100 \cdot \frac{1}{m_{\phi}^2} + 0.0000016 \cdot m_{\phi} - 0.00025 \;, \\
        c_{4,\epsilon,tW\phi}(m_{\phi}) &= - 0.0000032 \cdot m_{\phi} + 0.0036 \;, \\
        c_{5,\epsilon,tW\phi}(m_{\phi}) &= 1200 \cdot \frac{1}{m_{\phi}^2} - 0.0000026 \cdot m_{\phi} + 0.0021 \;, \\
        c_{6,\epsilon,tW\phi}(m_{\phi}) &= - 1200 \cdot \frac{1}{m_{\phi}^2} - 0.0000059 \cdot m_{\phi} + 0.0067 \;, \\
        c_{7,\epsilon,tW\phi}(m_{\phi}) &= 550 \cdot \frac{1}{m_{\phi}^2} - 0.0000047\cdot m_{\phi} + 0.0049\;.
    \end{split}
\end{align}
This gives the function for $\sigma_{\epsilon,tW\phi}$ as
\begin{align}
\begin{split}
    \sigma_{\epsilon,tW\phi}(c_t,c_{\tilde{t}}, c_V, m_{\phi}) = &c_{1,\epsilon,tW\phi}(m_{\phi}) \cdot c_V^2 c_t^2 + 
                                                         c_{3,\epsilon,tW\phi}(m_{\phi}) \cdot c_V c_t^3 + \\
                                                         &c_{4,\epsilon,tW\phi}(m_{\phi}) \cdot c_V c_t \tilde{c}_{t}^2 +
                                                         c_{5,\epsilon,tW\phi}(m_{\phi}) \cdot c_t^4 +  \\
                                                         &c_{6,\epsilon,tW\phi}(m_{\phi}) \cdot c_t^2 \tilde{c}_{t}^2 + 
                                                         c_{7,\epsilon,tW\phi}(m_{\phi}) \cdot \tilde{c}_{t}^4 \, .  
\end{split}
\end{align}
The fit functions for the total cross section $\sigma_{\text{tot},tW\phi}$ are given by
\begin{align}
    \begin{split}
        c_{1,\text{tot},tW\phi}(m_{\phi}) &= 940000 \cdot \frac{1}{m_{\phi}^2} + 0.0011 \cdot m_{\phi} - 1.8 , \\
        c_{2,\text{tot},tW\phi}(m_{\phi}) &= 0 ,\\
        c_{3,\text{tot},tW\phi}(m_{\phi}) &= - 2100000 \cdot \frac{1}{m_{\phi}^2} - 0.0015 \cdot m_{\phi} - 3 , \\
        c_{4,\text{tot},tW\phi}(m_{\phi}) &= 12000 \cdot \frac{1}{m_{\phi}^2} + 0.000055 \cdot m_{\phi} - 0.069 , \\
        c_{5,\text{tot},tW\phi}(m_{\phi}) &= 1300000 \cdot \frac{1}{m_{\phi}^2} + 0.000093 \cdot m_{\phi} - 0.84 , \\
        c_{6,\text{tot},tW\phi}(m_{\phi}) &= - 8600 \cdot \frac{1}{m_{\phi}^2} - 0.000035 \cdot m_{\phi} + 0.044 , \\
        c_{7,\text{tot},tW\phi}(m_{\phi}) &= 1300000 \cdot \frac{1}{m_{\phi}^2} - 0.0000043 \cdot m_{\phi} - 0.69.
    \end{split}
\end{align}
This gives the function for $\sigma_{\text{tot},tW\phi}$
\begin{align}
\begin{split}
    \sigma_{\text{tot},tW\phi}(c_t,c_{\tilde{t}}, c_V, m_{\phi}) = &c_{1,\text{tot},tW\phi}(m_{\phi}) \cdot c_V^2 c_t^2 + 
                                                         c_{tW\phi}(m_{\phi}) \cdot c_V c_t^3 + \\
                                                         &c_{4,\text{tot},tW\phi}(m_{\phi}) \cdot c_V c_t \tilde{c}_{t}^2 +
                                                         c_{5,\text{tot},tW\phi}(m_{\phi}) \cdot c_t^4 +  \\
                                                         &c_{6,\text{tot},tW\phi}(m_{\phi}) \cdot c_t^2 \tilde{c}_{t}^2 + 
                                                         c_{7,\text{tot},tW\phi}(m_{\phi}) \cdot \tilde{c}_{t}^4.  
\end{split}
\end{align}




\subsubsection{\texorpdfstring{$t\phi$}{tphi} production}

The fit functions for the coefficients in $\sigma_{\epsilon,t\phi}$ are given in the form $c_{i,\epsilon, t\phi}$ by
\begin{align}
    \begin{split}
        c_{5,\epsilon,t\phi}(m_{\phi}) &= 604.51 \cdot \frac{1}{m_{\phi}^2} - 0.00000053 \cdot m_{\phi} + 0.00032 ,\\
        c_{7,\epsilon,t\phi}(m_{\phi}) &= 150.39 \cdot \frac{1}{m_{\phi}^2} + 0.0000017 \cdot m_{\phi} - 0.0017.
    \end{split}
\end{align}
All other coefficients $c_i$ are zero. This gives the function for $\sigma_{\epsilon,ttH}$ as
\begin{equation}
    \sigma_{\epsilon,t\phi}(c_t,c_{\tilde{t}}, m_{\phi}) = c_{5,\epsilon,t\phi}(m_{\phi}) \cdot c_t^4 +  c_{7,\epsilon,t\phi}(m_{\phi}) \cdot \tilde{c}_{t}^4 .
\end{equation}
The fit functions for the total cross section $\sigma_{\text{tot},t\phi}$ are given by 
\begin{align}
    \begin{split}
        c_{5,\text{tot},t\phi}(m_{\phi}) &= 2948263.1 \cdot \frac{1}{m_{\phi}^2} + 0.0041 \cdot m_{\phi} - 6.35 ,\\ 
        c_{7,\text{tot},t\phi}(m_{\phi}) &= 1741654.4 \cdot \frac{1}{m_{\phi}^2} + 0.0014 \cdot m_{\phi} - 2.57.
    \end{split}
\end{align}
\GW{This} gives the function for $\sigma_{\text{tot},t\phi}$
\begin{equation}
    \sigma_{\text{tot},t\phi}(c_t,c_{\tilde{t}}, m_{\phi}) =  c_{5,\text{tot},t\phi}(m_{\phi}) \cdot c_t^4 +  c_{7,\text{tot},t\phi}(m_{\phi}) \cdot \tilde{c}_{t}^4 .
\end{equation}



\subsection{Implementation in \HiBo}

In this section, we will document the implementation of the four-top analysis in the code \HiBo. In \HiBo, we are able to implement a limit that is derived 
from coupling dependent acceptances. The four-top analysis \GW{presented} in \SH{\refse{sec:multi_top}} is the first analysis which is implemented in this way. Therefore, we here document the necessary code. For simplicity we will only
show an example for the implementation of the $ttH$ process. The limits for $tWH$ and $tH$ production are implement in the same way.

We start by importing some dependencies. These include some well-known \Code{Python} packages for data handling (e.g.\ \Code{pandas}) and \HiPr, as well as \HiBo. We also need to import some functions from \Code{ImplementationUtils}, which is part of the \HiTo installation.

\begin{minted}[bgcolor=bg]{python}
import pandas as pd
import numpy as np
from Higgs.tools.ImplementationUtils import (
    implementChannelLimit,
    implementChannelWidthLimit,
    fromHB5Table1,
    readHEPDataCsv,
)
from Higgs import Predictions
from Higgs import predictions as HP
from Higgs import bounds as HB
from Higgs.tools.LimitValidation import validateChannelLimit, 
validateChannelWidthLimit
import os, sys
import matplotlib.pyplot as plt
import pwlf
from scipy.interpolate import interp1d

sys.path.insert(0, os.path.dirname(os.path.dirname(os.getcwd())))
import MassResolutions as resolution
\end{minted}

We can calculate the upper limit on the signal events $n_s$ with \Code{MadAnalysis}. For this we need the observed and expected number of events in our most sensitive 
signal region. We then implement the expected and observed limit on $n_s$ in 
\GW{the considered}
mass range of $[350,1000]$ GeV. We also have to define the fit functions for the individual
coefficients of the total cross-section $\sigma_{tot}$ and cross-section times efficiency $\sigma \cdot \epsilon$. The number of signal events is defined as

\begin{equation}
    n_s = \epsilon_s \cdot \mathcal{L} \cdot \sigma.
\end{equation}

For Signal Region 8 of the CMS analysis the upper limit on the expected and observed number of signal are given by $n_{s,\mathrm{exp}}=6.94$ and $n_{s,\mathrm{obs}}=6.43$, \GW{respectively}. We  use this to implement the observed and expected limit on 

\begin{equation}
    \sigma \cdot\epsilon = \frac{n_s}{\mathcal{L}}.
\end{equation}

One also has to specifically define the production mode as \Code{["Htt"]} in \Code{\GW{HiggsBounds}}.

\begin{minted}[bgcolor=bg]{python}
df = pd.DataFrame(
    columns={
        "m",
        "obs",
        "exp",
    }
)

# fit functions for efficiency eff = sigma_s/sigma_tot 

xtot_c5 = lambda m: 2703921.1 * 1/m**2 + 0.0019 * m - 3.71
xtot_c7 = lambda m: 4300877.59 * 1/m**2 + 0.0048 * m - 8.14

xeff_c5 = lambda m: 2523.55 * 1/m**2 + -0.0000079 * m + 0.0076
xeff_c7 = lambda m: 6087.51 * 1/m**2 + 0.00000027 * m - 0.0039

prods = ["Htt"]

masses = np.linspace(350, 1000, 100)

df.m = masses
df.obs = [6.43/(1000*137)] * len(masses) 
df["exp"] = [6.94/(1000*137)] * len(masses)
\end{minted}

Now we can start with the actual limit implementation of the coupling-dependent acceptances. For this, we define the dependence of the acceptances on the individual couplings by our fit functions for the coefficients of $\sigma \cdot \epsilon$. This is done in the code block \Code{acceptances=[...]}. Here we assign the fit functions to the individual couplings defined by \Code{effCPeTopYuk} for the $\mathcal{CP}$-even and \Code{effCPoTopYuk} for the $\mathcal{CP}$-odd top-Yukawa coupling. The number (in our case four) depicts to which power the coupling contributes.

\begin{minted}[bgcolor=bg]{python}
df1 = pd.DataFrame(
    columns={
        "m",
        "AccCPe",
        "AccCPo"
    }
)

df1.m = masses
df1.AccCPe = xeff_c5(df1.m)
df1.AccCPo = xeff_c7(df1.m)

limitFile = implementChannelLimit(
    "1908.06463",
    {"channels": [[p, "tt"] for p in prods]},
    "https://arxiv.org/pdf/1908.06463.pdf",
    df,
    massResolution=resolution.tt["tttt"],
    acceptances=[
        {
            "couplingDepAcceptance": [
            [
                {"effCPeTopYuk": 4},
                {
                    "massDepAcceptance": df1.AccCPe.to_list(),
                    "massGrid": df1.m.to_list()
                }
            ],
           [
               {"effCPoTopYuk": 4},
               {
                   "massDepAcceptance": df1.AccCPo.to_list(),
                   "massGrid": df1.m.to_list()
               }
           ]
        ],
        "denominator": [
            [
                {"effCPeTopYuk": 4},
                {
                    "massDepAcceptance": df1.AccCPetot.to_list(),
                    "massGrid": df1.m.to_list()
                }
            ],
           [
               {"effCPoTopYuk": 4},
               {
                   "massDepAcceptance": df1.AccCPotot.to_list(),
                   "massGrid": df1.m.to_list()
               }
           ]
        ]
        }
        for p in prods
    ],
)
lim = HB.Limit(limitFile)

validateChannelLimit(
    lim
)
\end{minted}



\section{\texorpdfstring{$\kappa_{\lambda}$}{kappa\_lambda}-dependent Higgs pair production limits}
\label{app:kala_dep_limits}

In the following we provide an example of the definition of a cross-section limit that depends on the value of a particular coupling, in \SHd{particular} \SH{this case} the trilinear Higgs coupling. To stick to a concrete example, we provide the \texttt{json} file for the non-resonant $b\bar{b}\gamma\gamma$ ATLAS search~\cite{ATLAS:2025hhd}. 

We set the observed/expected limit to the value given by the experimental results at $\kappa_{\lambda}=1$. In this case 
\GW{those values are}
$\mu_{\mathrm{obs}}=3.8/\mu_{\mathrm{exp}}=3.7$, therefore we provide the following input in the \texttt{json} file: \texttt{observed} = $3.8 \cdot\sigma_{\mathrm{SM}}\cdot 2\cdot\mathrm{BR}(h_{125}\to b\bar{b})\cdot\mathrm{BR}(h_{125}\to \gamma\gamma)$ and \texttt{expected} = $3.7 \cdot\sigma_{\mathrm{SM}}\cdot 2\cdot\mathrm{BR}(h_{125}\to b\bar{b})\cdot\mathrm{BR}(h_{125}\to \gamma\gamma)$.

We then apply a multiplicative acceptance factor, $a$, to account for $\kappa_{\lambda}$-dependent limits: $\sigma_{\mathrm{obs}}(\kappa_{\lambda}) = a\cdot\sigma_{\mathrm{obs}}(\kappa_{\lambda}=1)$. In order to define $a$ we fit experimental data to a polynomial function of the form of \cref{eq:noresacc}. 
\GW{If} there is hep data available we simply fit to experimental bounds. In this case, however, only exclusion limits for $\kappa_{\lambda}$ were provided. We therefore fit only three values: 
\GW{(i)} for the lower bound $\kappa_{\lambda}=-1.7$ we set $\sigma_{\mathrm{SM}}(\kappa_{\lambda}=-1.7)\cdot 2\cdot\mathrm{BR}(h_{125}\to b\bar{b})\cdot\mathrm{BR}(h_{125}\to \gamma\gamma)$, where $\sigma_{\mathrm{SM}}(\kappa_{\lambda}=-1.7)$ is computed with \HiPr and yields 212.14 fb, 
\GW{(ii)} the experimental upper bound $3.8\cdot\sigma_{\mathrm{SM}}(\kappa_{\lambda}=1)\cdot 2\cdot\mathrm{BR}(h_{125}\to b\bar{b})\cdot\mathrm{BR}(h_{125}\to \gamma\gamma)$ and 
\GW{(iii)} $\sigma_{\mathrm{SM}}(\kappa_{\lambda}=6.6)\cdot 2\cdot\mathrm{BR}(h_{125}\to b\bar{b})\cdot\mathrm{BR}(h_{125}\to \gamma\gamma)$, where $\sigma_{\mathrm{SM}}(\kappa_{\lambda}=6.6)=267.61$ fb.

The coefficients $A, B$ and $C$ in \cref{eq:noresacc} are implemented in 
\texttt{couplingDepAcceptance} in the snippet below for the corresponding power of $\kappa_{\lambda}$, which is specified by \texttt{effLam}, and the coefficients $E,F$ and $G$ are specified in the \texttt{denominator} section. We also note here that even though also
data from 13.6 TeV was used in the implementation of this limit, we use the cross sections at 13 TeV because of the need to select one collider option to apply the limits in \HiTo.

\begin{minted}[bgcolor=bg]{python}
{
    "limitClass": "PairProductionLimit",
    "id": 250703495,
    "reference": "2507.03495",
    "source": "",
    "citeKey": "ATLAS:2025hhd",
    "collider": "LHC13",
    "experiment": "ATLAS",
    "luminosity": 308,
    "process": {
        "firstDecay": [
            "gamgam"
        ],
        "secondDecay": [
            "bb"
        ]
    },
    "analysis": {
        "equalParticleMasses": true,
        "massResolution": {
            "firstParticle": {
                "absolute": 0,
                "relative": 0.015
            },
            "secondParticle": {
                "absolute": 40,
                "relative": 0
            }
        },
        "acceptances": [
            {
            "denominator": [
                [
                    {
                        "effLam": 0
                    },
                    {
                        "constantAcceptance": 8.3926
                    }
                ],
                [
                    {
                        "effLam": 1
                    },
                    {
                        "constantAcceptance": -1.42855
                    }
                ],
                [
                    {
                        "effLam": 2
                    },
                    {
                        "constantAcceptance": 2.68485
                    }                       
                ]
            ],
            "couplingDepAcceptance": [
                [
                    {
                        "effLam": 0
                    },
                    {
                        "constantAcceptance": 8.3926
                    }
                ],
                [
                    {
                        "effLam": 1
                    },
                    {
                        "constantAcceptance": 0.256298
                    }
                ],
                [
                    {
                        "effLam": 2
                    },
                    {
                        "constantAcceptance": 1
                    }                       
                ]
            ]
            }
        ],
        "grid": {
            "massFirstParticle": [
                125.0
            ]
        },
        "limit": {
            "observed": [
                0.00032652880782958976
            ],
            "expected": [
                0.0003179359444656533
            ]
        }
    }
}
\end{minted}